\documentclass[fleqn,usenatbib]{mnras}

\usepackage{newtxtext,newtxmath, xspace, xcolor}
\usepackage{amsmath}
\usepackage{caption}
\usepackage{subcaption}
\usepackage[normalem]{ulem}

\usepackage{hyperref}
\usepackage[T1]{fontenc}

\newcommand{\Balrog}{\textsc{Balrog}\xspace}
\newcommand{\mdet}{\textsc{Metadetection}\xspace}
\newcommand{\mcal}{\textsc{Metacalibration}\xspace}
\newcommand{\vecc}{{\bf c}}
\newcommand{\vecu}{{\bf u}}
\newcommand{\vecn}{{\bf n}}
\newcommand{\vecm}{{\bf m}}
\newcommand{\vecz}{{\bf z}}
\newcommand{\matE}{E}
\newcommand{\matD}{D}
\newcommand{\likeli}{\mathcal{L}}

\DeclareRobustCommand{\VAN}[3]{#2}
\let\VANthebibliography\thebibliography
\def\thebibliography{\DeclareRobustCommand{\VAN}[3]{##3}\VANthebibliography}

\usepackage{graphicx}	
\usepackage{amsmath}	
\usepackage{enumitem}  
\usepackage{float}
\restylefloat{table}
\usepackage{eso-pic}

\AddToShipoutPictureBG*{%
  \AtPageUpperLeft{%
    \hspace{0.75\paperwidth}%
    \raisebox{-5\baselineskip}{%
      \makebox[0pt][l]{\textnormal{DES-2025-0948}}
}}}%

\AddToShipoutPictureBG*{%
  \AtPageUpperLeft{%
    \hspace{0.75\paperwidth}%
    \raisebox{-6\baselineskip}{%
      \makebox[0pt][l]{\textnormal{FERMILAB-PUB-25-0751-PPD}}
}}}%

\title[DES Y6 Source Redshift Calibration]{Dark Energy Survey Year 6 Results: Redshift Calibration of the Weak Lensing Source Galaxies}

\author[Yin et al.]{
\parbox{\textwidth}{
\Large B.~Yin$^{1}$\thanks{E-mail: boyan.yin@duke.edu}, 
A.~Amon,$^{2}$
A.~Campos,$^{3}$
M.~A.~Troxel,$^{1}$
W.~d'Assignies,$^{4}$
G.~M.~Bernstein,$^{5}$
G.~Camacho-Ciurana,$^{6}$
S.~Mau,$^{1,7,8}$
M.~R.~Becker,$^{9}$
G.~Giannini$^{10,11}$ 
A.~Alarcon,$^{6}$
D.~Gruen,$^{12}$
J.~McCullough,$^{2}$
M.~Yamamoto$^{1,2}$
D.~Anbajagane,$^{11}$
S.~Dodelson,$^{10,11,13}$
C.~S{\'a}nchez,$^{5}$
J.~Myles,$^{2}$
J.~Prat,$^{10,14}$
C.~Chang,$^{10,11}$
M.~Crocce,$^{4,15}$
K.~Bechtol,$^{16}$
A.~Fert{\'e},$^{7,8,17}$
M.~Gatti,$^{11}$
N.~MacCrann,$^{18}$
R.~Marco,$^{19}$
A.~Porredon,$^{20}$ 
D.~Sanchez Cid,$^{20,21}$
T.~Schutt,$^{7,8}$
M.~Tabbut$^{11}$
C.~To,$^{10,11}$
T.~Abbott,$^{22}$
M.~Aguena,$^{23}$
O.~Alves,$^{24}$
D.~Bacon,$^{25}$
J.~Blazek,$^{26}$
S.~Bocquet,$^{12}$
D.~Brooks,$^{27}$
R.~Camilleri,$^{28}$
A.~Carnero Rosell,$^{29,30,31}$
M.~Carrasco Kind,$^{32,33}$
J.~Carretero,$^{4}$
F.~Castander,$^{6,15}$
R.~Cawthon,$^{34}$
C.~Conselice,$^{35,36}$
L.~da Costa,$^{30}$
M.~da Silva Pereira,$^{37}$
T.~Davis,$^{28}$
J.~De Vicente,$^{20}$
S.~Desai,$^{38}$
H.~Diehl,$^{13}$
C.~Doux,$^{5,39}$
A.~Drlica-Wagner,$^{10,11,13}$
T.~Eifler,$^{40,41}$
J.~Elvin-Poole,$^{42}$
S.~Everett,$^{43}$
B.~Flaugher,$^{13}$
P.~Fosalba,$^{6,15}$
D.~Francis de Souza,$^{41}$
J.~Frieman,$^{10,11,13}$
J.~Garcia-Bellido,$^{44}$
E.~Gaztanaga,$^{6,15,25}$
P.~Giles,$^{45}$
G.~Gutierrez,$^{13}$
S.~Hinton,$^{28}$
D.~Hollowood,$^{46}$
K.~Honscheid,$^{47,48}$
D.~Huterer,$^{38}$
B.~Jain,$^{5}$
D.~James,$^{49}$
K.~Kuehn,$^{50,51}$
S.~Lee,$^{41}$
H.~Lin,$^{13}$
J.~Marshall,$^{52}$
J.~Mena-Fern\'{a}ndez,$^{39}$
F.~Menanteau,$^{32,33}$
R.~Miquel,$^{4,53}$
J.~Muir,$^{54,55}$
L.~Ofer,$^{27}$
R.~Ogando,$^{56,57}$
A.~Palmese,$^{3}$
D.~Petravick,$^{33}$
A.~Plazas Malag{\'o}n,$^{7,17}$
A.~Roodman,$^{7,17}$
R.~Rosenfeld,$^{30,58}$
S.~Samuroff,$^{4,26}$
E.~Sanchez,$^{20}$
I.~Sevilla,$^{20}$
E.~Sheldon,$^{59}$
T.~Shin,$^{57}$
M.~Smith,$^{60}$
E.~Suchyta,$^{61}$
M.~Swanson,$^{3}$
G.~Tarle,$^{24}$
D.~Thomas,$^{13}$
V.~Vikram,$^{5}$ 
A.~Walker,$^{22}$ 
and
P.~Wiseman$^{62}$
\begin{center} (DES Collaboration) \end{center}
}
}

\date{Accepted XXX. Received YYY; in original form ZZZ}

\pubyear{\the\year{}}

\begin{document}
\label{firstpage}
\pagerange{\pageref{firstpage}--\pageref{lastpage}}
\maketitle

\begin{abstract}
Determining the distribution of redshifts for galaxies in wide-field photometric surveys is essential for robust cosmological studies of weak gravitational lensing. We present the methodology, calibrated redshift distributions, and uncertainties of the final Dark Energy Survey Year 6 (Y6) weak lensing galaxy data, divided into four redshift bins centered at $\langle z \rangle = [0.414, 0.538, 0.846, 1.157]$. We combine independent information from two methods on the full shape of redshift distributions: optical and near-infrared photometry within an improved Self-Organizing Map $p(z)$ (SOMPZ) framework, and cross-correlations with spectroscopic galaxy clustering measurements (WZ), which we demonstrate to be consistent both in terms of the redshift calibration itself and in terms of resulting cosmological constraints within 0.1$\sigma$. We describe the process used to produce an ensemble of redshift distributions that account for several known sources of uncertainty. Among these, imperfection in the calibration sample due to the lack of faint, representative spectra is the dominant factor. The final uncertainty on mean redshift in each bin is $\sigma_{\langle z\rangle} = [0.012, 0.008,0.009, 0.024]$. We ensure the robustness of the redshift distributions by leveraging new image simulations and a cross-check with galaxy shape information via the shear ratio (SR) method.

\end{abstract}


\section{Introduction}
The use of weak lensing data for cosmological inference, since its initial measurement in \citet*{Bacon_2000}, \citet*{Kaiser_2000}, \citet{VanWaerbke_2000} and \citet{Wittmann_2000}, has become a robust probe of large-scale structure. Cosmic shear, which is the coherent distortion of apparent galaxy shapes in response to the foreground mass distribution through gravitational lensing, provides a statistical measure. When combined with galaxy-galaxy lensing and galaxy clustering in a 3$\times$2pt analysis, it has strong constraining power on the structure growth parameter $S_8$ by breaking parameter degeneracies. Further separating the galaxies into multiple (so-called tomographic) redshift bins, the expansion and structure growth history of the Universe is being traced, constraining the dark energy equation of state parameters $w_0$ and $w_a$, especially in a 3$\times$2pt analysis where lens bins provide additional information at different redshift regimes. Stage III weak lensing surveys including KiDS, HSC, and DES \citep[e.g., most recently analysed in ][]{kids_3x2pt, HSCY3_3x2pt, DESY3_3x2pt} have demonstrated the power of cosmic shear and 3$\times$2pt analyses. Looking to the near future, Stage IV surveys including Euclid, LSST, and Roman \citep{Euclid, LSST_SRD, Roman_SRD} are also optimized to use them as some of their key probes.

In gravitational lensing, the known distance to the lensed object is required for interpreting its shape distortion physically. Cosmic shear is therefore highly dependent on accurate knowledge of the distance, or redshift, of the ensemble of source galaxies. This is not an easy process: the observed colour is degenerate under joint transformations of galaxy redshift and type, that is, galaxies at different redshifts can appear similar in broadband colour and flux due to their intrinsic properties, as illustrated in Figure~1 in \citet{Buchs_SOMPZ}. Breaking this degeneracy is commonly impossible given photometry in only four broad bands ($griz$), as available in DES Y6.

People have devised several methods in response to this challenge (for a detailed review see \citet*{newman_gruen}). In 2019, \citet{Buchs_SOMPZ} developed SOMPZ, which is a statistical method that uses multi-band deep fields -- most successfully when combining optical + NIR data -- to break the colour-redshift degeneracy, spectroscopic or high-quality multi-band photometric samples to provide redshift information, and the machine learning method Self-Organizing Map (SOM) to group similar galaxies together. The method is fully data driven, relying on calibration samples rather than models for the galaxy spectral energy distribution (SED), as opposed to template-fitting. Methods similar to SOMPZ have been adopted successfully in KiDS \citep{KiDSlegacy_redshift} and DES \citep*{y3-wlpz}. In this paper, we use the SOMPZ framework of DES Y3, and incorporate an improved SOM algorithm that is more robust for faint galaxies \citep{y3-highz,somf}. We also include $g$ in addition to $riz$ band measurements in the Y6 analysis, enabled by the improvement of chromatic PSF modeling in DES Y6 \citep{desy6_psf}. Challenges remain, particularly when compiling the redshift calibration sample: relying only on spectroscopic data would introduce incompleteness, especially at faint magnitudes and high redshift. To ensure a complete sample, we incorporate photometric redshifts from \texttt{PAUS} \citep{paus-cosmos} and \texttt{COSMOS2020} \citep{cosmos2020}, but they suffer potential bias in the high redshift and faint apparent magnitude galaxy populations. In stage IV, such limitations may be better resolved by dedicated redshift calibration follow-up surveys with well-understood selection functions, such as Subaru-PFS for Roman(SuPR) \citep{supR2025}, Complete Calibration of the colour-Redshift Relation Survey (C3R2) \citep{c3r2_2017,c3r2_2019,c3r2_2021},  DESI C3R2 (DC3R2) \citep{dc3r2}, and 4MOST C3R2 (4C3R2) \citep{4c3r2}. 

DES has developed a pipeline to constrain redshift calibration uncertainties by combining methods based on complementary observables. We use the position information of our sheared galaxies to cross-correlate with the positions of a wide area spectroscopic sample on which we have no completeness requirements, and constrain the SOMPZ-predicted redshift distribution using this clustering redshift (WZ) measurement \citep{y6-wz}. We use the shape information of our sheared galaxies, cross-correlated with the position of our lens galaxies at small angular scales (unused in 3$\times$2pt analysis), and construct the shear ratio (SR) as a cross-check of our redshift results \citep{y6-ggl}.  WZ and SR are independent of the colour-redshift degeneracies and are not affected by the calibration sample uncertainties in SOMPZ, making them valuable to constrain SOMPZ redshift uncertainty and cross-check our redshift calibration results. 

This is one of the four publications that describe the redshift calibration for the DES Y6 3$\times$2pt analysis: 1) \citet{y6-clpz} describes the redshift calibration of the lens galaxy sample; 2) This paper describes the redshift calibration of the source galaxy sample; 3) \citet{y6-wz} describes the redshift calibration from clustering measurements; and 4) \citet{y6-modes} describes the redshift uncertainty sampling method used in the cosmological inference. All together, these studies describe the Y6 redshift calibration pipeline. In addition, two related analyses play a crucial role to this end: a) \citet*{y6-imagesims} describes the image simulations on which our methods are validated, and the corresponding source redshift correction due to galaxy blending along the line of sight; and b) \citet{y6-ggl} describes the shear ratio validation for the lens and source redshift distributions.

\begin{figure*}
    \centering
    \includegraphics[width=\textwidth]{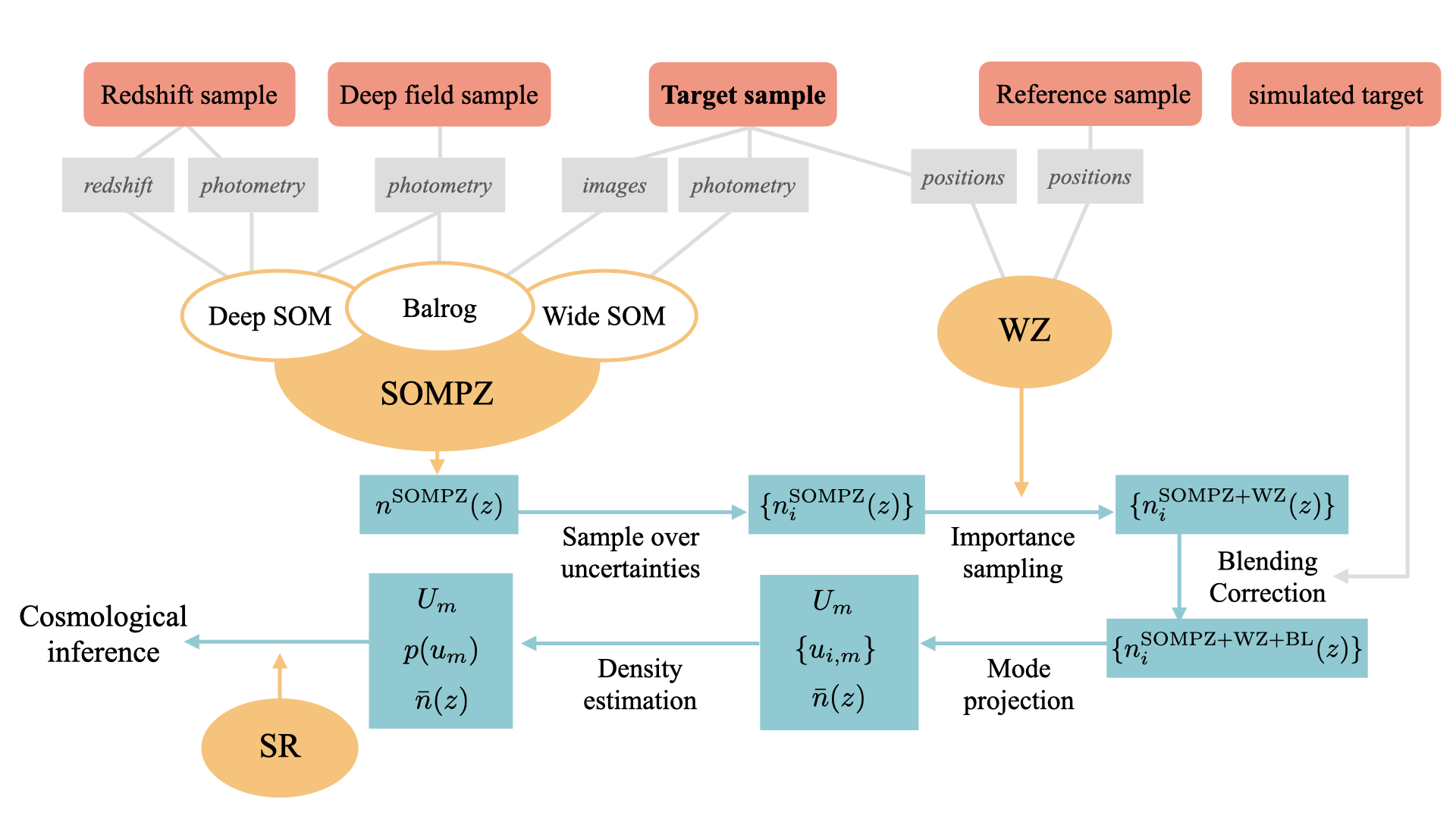}
    \caption{\label{fig:flowchart} Schematic overview of the DES Y6 source redshift calibration pipeline. The goal is to calibrate the redshift distribution of our target DES Y6 weak lensing source sample. Photometric measurements from the target sample are used within SOMPZ with the redshift sample, the multi-band deep field sample, and the \Balrog simulation linking the deep and wide field, to calibrate $n^{\rm{SOMPZ}}(z)$. Sampling over redshift uncertainties generates multiple realizations of this distribution ${n^{\rm{SOMPZ}}(z)}$, which are then constrained through importance sampling using spatial correlation with the reference sample (WZ). Blending effects are corrected using the simulated target sample, and the resulting realizations are projected onto modes using a cosmology-sensitive principal component analysis (PCA) method, before entering cosmological inference. We finally cross-check using the target sample shape information in the SR consistency test.}

\end{figure*}

This paper is structured as follows: We provide an overview of our photometric redshift calibration pipeline in Section \ref{sec:Overview}. Section \ref{sec:Data} describes the Y6 weak lensing source galaxy sample, the simulated samples, and the calibration samples used for redshift calibration. Section \ref{sec:SOMPZ} describes the SOMPZ method we use to calibrate the redshift distributions, while Section \ref{sec:redshift_calibration_uncertainty} describes the calibration and quantification of different sources of redshift uncertainty. In Section \ref{sec:redshift_calibration_pipeline}, SOMPZ results are combined with WZ. Section \ref{sec:blending} further corrects for the impact of blending using image simulations, and compresses redshift realizations using mode projections for cosmological likelihood analyses. Final results including consistency tests using SR and cosmological inference are presented in Section \ref{sec:results_and_discussion}, with further discussion in Section \ref{sec:discussion}.

\section{Overview of the methodology}
\label{sec:Overview}
The goal of this paper is to determine the joint probability distribution $p\left[n_1(z), n_2(z), n_3(z), n_4(z)\right]$ of the weighted redshift distributions for the four tomographic bins of DES Y6 weak lensing source galaxies, accumulated through six years of observations, in a form that can be efficiently sampled by the likelihood analysis. The full redshift calibration pipeline is shown in Fig.~\ref{fig:flowchart}: 

\begin{enumerate}[labelindent=0pt,
  itemindent=\parindent,
  listparindent=\parindent,
  leftmargin=0pt,
  align=left]
    \item \textbf{Creating the Self-Organizing Maps}: using source (photometric) galaxies and calibration galaxy samples from the redshift calibration sample, deep field photometric sample, and synthetic source sample. The preliminary redshift distributions are then determined using the SOM results, re-weighting the redshift calibration sample to represent the relative abundances of source galaxies, with an intermediate step utilizing deep field galaxies to break colour-redshift degeneracy. The process and the Y6 SOM result are shown in Section~\ref{sec:SOMPZ}.
    \item \textbf{Sampling uncertainty due to limitations of the calibration samples}: the limited area and number of the redshift calibration sample and deep field galaxies, the photometric zero-point calibration uncertainty among the four deep fields, and the photometric redshift bias in the calibration samples \texttt{COSMOS2020} and \texttt{PAUS}. The modeling and the propagation of these calibration sample uncertainties to the source redshift distribution are described in Section~\ref{sec:redshift_calibration_uncertainty}, with a discussion of the Y6 redshift uncertainty results. The process produces $10^8$ $n(z)$ realizations incorporating all the sources of uncertainty besides blending.
    \item \textbf{Compressing the redshift distributions}: The redshift realizations are passed through a linear filter that removes modes of variation that have little detectable influence on the summary statistics of the data, namely on the shear-shear correlation functions $\xi_{\pm}$ and the galaxy-galaxy lensing signal $\gamma_t.$  This "mode projection" method, described in Section~\ref{sec:importance} and \citet{y6-modes}, consists of deriving an encoding matrix $\matE$ and a decoding matrix $\matD$. The original vector $\vecn$ of $n(z)$ values is transformed into a much smaller vector $\vecu = \matE\vecn$, and then reforming as $\hat\vecn = \matD \vecu$. At this stage, the mode projection is serving as a filter to eliminate cosmologically irrelevant uncertainties from the $n(z)$ distribution. 
    \item \textbf{Combining with clustering redshifts}: For each mode-projected redshift realization $\hat\vecn$, we calculate its likelihood $\likeli(D_{\rm WZ} | \hat\vecn)$ of generating the observed small-scale angular correlations between source galaxies and eBOSS spectroscopic galaxies \citep{eboss_dawson}.  The measurement of this clustering redshift (WZ) data, and the construction of the likelihood function for the WZ data as a function of $n(z)$, are described in Section \ref{sec:WZ}.  The redshift realizations are then accepted or rejected in proportion to the WZ likelihood.  This importance-sampling process reduces the $10^8$ SOMPZ $n(z)$ realizations to $10^4$ $n(z)$ realizations drawn from the joint probability of matching both the SOMPZ and WZ data, marginalized over the SOMPZ calibration variables, as described in Section \ref{sec:importance}.
    \item \textbf{Applying a blending correction}: We further correct for the shear-dependent blending (BL) of the source galaxies using image simulations. 
    As redshift distributions for lensing are commonly weighted by their response of their measured shape to constant shear applied to the entire image (quantified for our purposes by \mdet), a galaxy with blended light from a foreground, lower-redshift galaxy will respond to shearing both at its own redshift (in a biased way) as well as at the redshift of its blending companion. This overall correction for blending can be computed with survey-specific simulations that shear galaxies in true redshift slices independently from one another. Posterior realizations of this correction are implemented on top of each SOMPZ+WZ realization to produce $8\times10^4$ $n(z)$ realizations. The process is detailed in Section~\ref{sec:bl_methodology} and \citet{y6-imagesims}, with results presented in \ref{sec:final_nz}.
    \item \textbf{Final smoothing and compression of the distributions}: The samples from the joint SOMPZ$+$WZ$+$BL posterior are then fed into the mode-projection algorithm again in Section~\ref{sec:modes}, yielding new matrices $\matE$ and $\matD$, and values of $\vecu$ for each sample. The compression of the $\vecn$ realizations to $M=7$ components of $\vecu$ chosen so that the discarded residuals $\vecn-\matD \vecu$ generate changes to the model data vector that correspond to an average $\langle \chi^2 \rangle < 0.1$ using the covariance matrix of the observables for Y6 data. 
    \item \textbf{Correlating redshift and shear-bias calibration parameters for inference}: The distribution of the $\vecu$ values for the SOMPZ$+$WZ$+$BL samples is, by design, consistent with independent unit-variance normal distributions for each mode $u_i$. However, the blending correction introduces correlation between shear multiplicative bias $\vecm$ and mean redshift shift $\Delta\vecz$, and propagates to $\vecm$-$\vecu$ correlation.  The representation of $p\left[n_1(z),\ldots,n_4(z)\right]$ that we use in cosmological investigations, therefore, is described by $\vecn = \matD \vecu$, where $\vecu$ and $\vecm$ are jointly sampled from the correlated $\vecm$-$\vecu$ prior. 

\end{enumerate}
\par\vfill

\section{Data}
\label{sec:Data}
This section introduces the four samples used in the SOMPZ method: our target sample for which we want to infer the redshift distributions -- the DES Y6 weak lensing sources (Section \ref{sec:DES_Y6_weak_lensing_source_sample}), the DES deep field data (Section \ref{sec:DES_Y6_deep_field}), the DES synthetic source sample (Section \ref{sec:DES_Y6_synthetic_source_sample}), and the redshift calibration sample (Section \ref{sec:Redshift_calibration_samples}). The area and number of galaxies in each sample are summarized in Table~\ref{tab:data_table}. Additionally, we describe the DES Y6 image simulation sample (Section \ref{sec:DES_Y6_image_simulations}), which is used to validate our methodology and correct for blending effects in the redshift calibration.

\begin{table}
\centering
\begin{tabular}{lcc}
& Sky Area  [$deg^2$]   &$\text{N}_\text{galaxies}$  \\
\hline
Redshift Sample         & 1.38 & 99,732 \\
Deep Sample            & 5.88 & 489,886 \\
\Balrog           & 4031 & 41,013,618 \\
Weak Lensing Source Sample  & 4031 & 139,662,173 \\
\hline
\end{tabular}
\caption{Sky area and number of galaxies for the four catalogs used in SOMPZ. The weak lensing source sample $\text{N}_\text{galaxies}$ in the table is selected from the $\sim$150M galaxies in \citeauthor*{y6-metadetect}(\citeyear{y6-metadetect}) using equation~(\ref{eq:redshift_cut}) and the source and lens sample joint mask.}
\label{tab:data_table}
\end{table}

\subsection{DES Y6 weak lensing source sample}
\label{sec:DES_Y6_weak_lensing_source_sample}
This work provides the calibrated redshift distributions for the DES Y6 weak lensing shape measurements. The source catalog, presented in \cite*{y6-metadetect}, is a subset of the DES Y6 Gold catalog of photometric objects \citep{y6-gold}, with selections to ensure robust shape measurement with \mdet. After the applied \mdet selections, it consists of $\sim$150M galaxies and covers 4422 deg$^2$ of the Southern sky with an effective number density of 8.22 galaxies per arcmin$^2$ and shape noise  $\sigma_e=0.29$. New colour-dependent PSF modeling presented in \cite{desy6_psf} resulted in improved control of our PSF models compared to Y3, particularly for the $g$-band, which enabled \mdet galaxy flux measurements over $griz$ (rather than only $riz$ in DES Y3) -- a significant gain for redshift calibration. Note that for the \mdet\ catalog, all objects are detected on $riz$ multi-band coadded images, and their per-band fluxes in $griz$ are measured on single-band coadded images by forced photometry \citep{y6-gold}. 

For robust redshift calibration and a more homogeneous photometric catalog, further selections are imposed on the weak lensing source sample. Specifically, bright magnitude limits are imposed, which removes nearby galaxies for which negligible lensing signal is expected. Furthermore, galaxies with unphysical colours are removed, as they are
assumed to be caused by catastrophic flux measurement failures, transient sources, or artifacts. The final weak lensing photometric selection is: 
\begin{eqnarray} \label{eq:redshift_cut}
&g, r, z > 15.2 \nonumber  \\
&18.2 < i < 24.7,  \\
&-1.5 < g-r, r-i, i-z < 4. \nonumber
\end{eqnarray}
The $i$-band faint limit we implement is roughly 1 magnitude deeper than in Y3. This $i$-band faint boundary is chosen to minimize the outlier fraction rate for the \textsc{COSMOS} narrow-band redshifts, which is the primary calibration catalog at faint magnitudes. In DES Y3, the $i$-band limit was set to 23.5 for using \textsc{COSMOS2015}, corresponding to the mid-point of [23,24] validation range, where the normalized median absolute deviation $\sigma = 0.022$ and outlier rate $\eta = 6.7$ in \citet{cosmos2015}. In DES Y6, with \textsc{COSMOS2020}, we extend the limit to 24.5 (the mid-point of [24,25] validation range) which has a similar deviation $\sigma = 0.021$ and lower outlier rate $\eta = 5.9$ in \texttt{FARMER-LePhare} of \citet{cosmos2020}. \textsc{COSMOS2020} uses the HSC $i$-band, so the 24.5 limit is shifted to 24.7 (+0.2 mag) to match the Gaussian aperture flux measured in DES \mdet. This offset is measured by converting the COSMOS $i$-band magnitude to DES Y6 $i$-band magnitude measured with pre-PSF moments in Fourier space. The same +0.2 magnitude adjustment is applied to the bright limit in $griz$ bands. The selections on colour are equivalent to those in the Y3 analysis \citep*{y3-wlpz}. Finally, the mask produced jointly from the shear catalog and magnitude-limited lens sample \citep[see][]{y6-mask} is applied.

A practical difference compared to the Y3 \mcal shape catalog \citep{y6-metadetect} is the computation of the "shear response", used to calibrate the raw galaxy ellipticity measurement. As \mdet\ does not provide shear responses for individual objects, the shear responses must be averaged over sets of galaxies meeting a certain selection criterion that is repeated on both sheared and unsheared images, with imperfect selections inducing a bias in the resulting shear. For this we use the main factors for variations in shear response cells -- noise level and resolvedness -- by considering cells $k$ in a two-dimensional grid of object size ratio $\text{T}_\text{ratio}$ and signal to noise $S/N$, such that
\begin{equation}\label{eqn:responseij}
\langle R_{ij}^k \rangle=  \frac{ \langle e_i^{k,+} \rangle - \langle e_i^{k,-} \rangle}{\Delta \gamma_j} \, .
\end{equation}
This \mdet response includes both the shear and selection response, which are computed separately in Y3 using \mcal, as well as selection effects related to redshift bin assignment. We take the mean response as the average of the diagonal elements in the response matrix,
\begin{equation}\label{eqn:response}
\langle R^k \rangle \equiv \frac{\langle R_{11}^k \rangle + \langle R_{22}^k \rangle}{2} \, .
\end{equation}
Our statistical weight definition for shear is the same as in Y3, using the \mdet shear response, and the measured ellipticity variance $\sigma_{e}^{k}$ in each grid cell $k$,
\begin{equation}\label{eqn:statistical_weight}
w_{\text{stat}}^k = (\sigma_{e}^{k})^{-2} \langle R^k \rangle^2.
\end{equation}

\begin{figure*}
    \centering
    \includegraphics[width=0.7\textwidth]{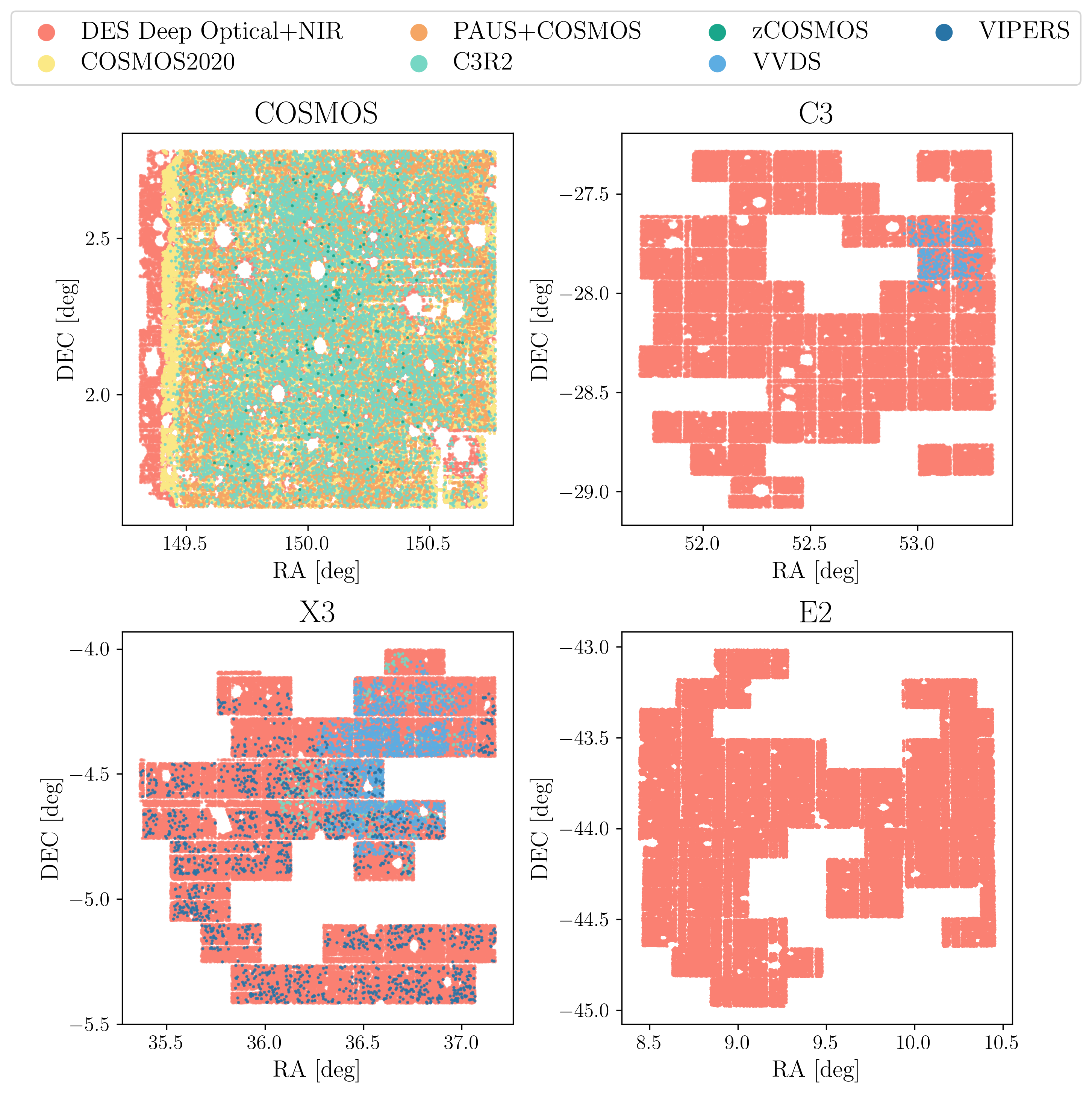}
    \caption{\label{fig:deep field} The four DES optical+NIR deep fields used for the SOMPZ redshift analysis, with joint photometry from deep DES $ugriz$ and VIDEO/UltraVISTA ${JHK_s}$ bands (red). The overlapping redshift calibration sample is shown from narrow-band \texttt{COSMOS2020} (yellow) and \texttt{PAUS+COSMOS} (orange), as well as spectroscopy from C3R2, zCOSMOS, VVDS and VIPERS (green to blue). The holes in the image are due to masking of artifacts. The total area and galaxy number after masking are shown numerically in Table~\ref{tab:data_table}.}
\end{figure*}

\begin{figure*}
    \centering
    \includegraphics[width=\textwidth]{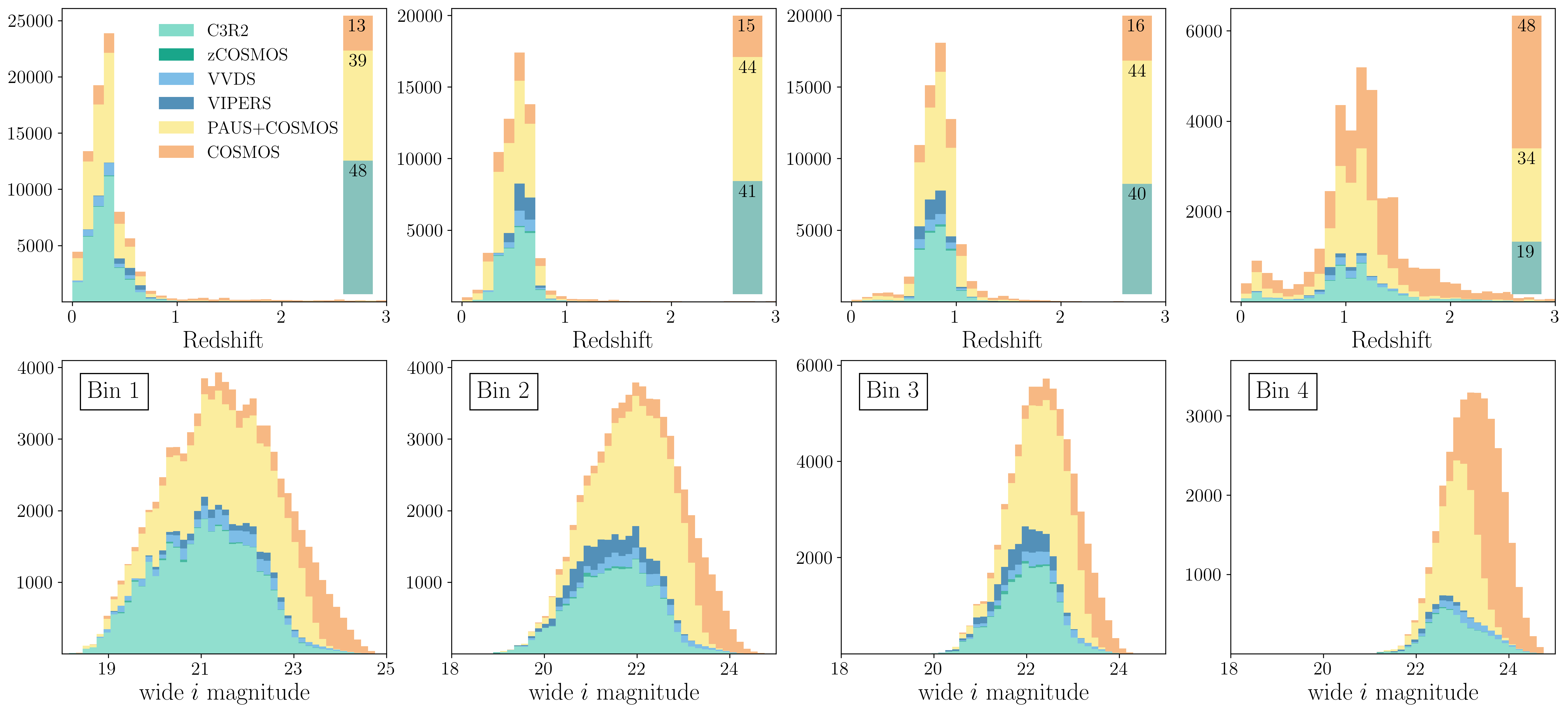}
    \caption{\label{fig:zsamples hist} The distribution of redshift calibration data that informs each redshift bin, shown as a function of redshift (top panel) and DES $i-$band magnitude (bottom panel). Each galaxy in this histogram is weighted by the combination of the shear response and statistical weight, and the \Balrog injection count. The bar charts (inset) show the relative contributions from the different redshift calibration samples, where we preferentially use spectroscopic redshifts (green), then consider photometric narrow-band \texttt{PAUS} (yellow) and \texttt{COSMOS2020} (orange), to ensure completeness. The calibration of the high redshift bin, comprised of the most faint galaxies, is dominated by photometric narrow-band data.}
\end{figure*}
\subsection{DES Y6 optical+NIR deep field data}
\label{sec:DES_Y6_deep_field}

The DES deep field data is comprised of the DES $ugriz$ bands and $\rm{JHK_s}$ UltraVISTA + VIDEO near-infrared (NIR) bands \citep{McCracken12,Jarvis13}. This joint eight-band photometry is crucial for breaking the colour-redshift degeneracy. In DES Y6, we use an updated version of the \citet{y3-deep} catalog, which spans four deep fields: C3, X3, E2, and COSMOS, with areas of 1.38, 1.94, 3.29, 3.32 square degrees respectively, as shown in Fig.~\ref{fig:deep field}.  The limiting (BDF, bulge + disk model) magnitude for all $ugriz\rm{JHK_s}$ bands is 25.4 -- much fainter than the limit of our source galaxy sample; while the photometric calibration uncertainty in each of the $ugriz\rm{JHKs}$ bands is: 0.055, 0.005, 0.005, 0.005, 0.005, 0.008, 0.008, 0.008, respectively, in magnitudes. The Y6 deep field catalog consists of the same objects also detected in the Y3 version \citep{y3-deep}, but with re-computed photometry using an updated fitting algorithm (\textsc{fitvd}) that matches the wide field algorithm \citep{y6-balrog}. Following Y3, we require each deep-field galaxy to have at least one wide-field detection in the DES Y6 synthetic source sample, limiting calibrators to those relevant for our target sample.

\subsection{DES Y6 synthetic source sample}
\label{sec:DES_Y6_synthetic_source_sample}

A necessary quantity for the SOMPZ method is the distribution of wide-field measured photometry given the deep-field (comparatively noiseless, true) photometry, $P(m_{\rm meas} | m_{\rm true})$. This distribution, which connects the deep and wide fields information, is also referred to as the "transfer function". In DES Y6, we measure this through \Balrog, a suite of synthetic source injections (SSI, \citealt{y6-balrog}). \Balrog injects a large catalog of galaxy and star model images into DES DECam images -- containing real sources, as well as the actual noise, sky-background, and other systematics -- and post-processes the modified images using the same pipeline as the real data. The output is a catalog of synthetic source measurements, for which we in addition know the true, injected properties. This can then be analyzed to extract variants of the distribution, $P(m_{\rm meas} | m_{\rm true})$. The light profiles of the synthetic sources are drawn from models fit to objects in the DES Y6 deep fields. These fits serve as our "truth catalog", given that they have significantly smaller uncertainties than the wide field measurements.

The DES Y6 \Balrog includes two advances that improve the robustness of the SOMPZ methodology. First, we have performed synthetic source injection (SSI) over the entire DES footprint, which accurately samples all the observing conditions and other systematics of the survey. The second is an optimized weighting scheme wherein a subset of the injections preferentially contains galaxies that are bright enough to likely pass our weak lensing selection cuts. Another subset further optimizes the injection for redshift analyses in particular by injecting sources only if they also have redshift estimates. See Section 3.2 in \citet{y6-balrog} for more details on the optimized injection scheme. Overall, these updates provide a factor $\mathcal{O}(10)$ more sources of relevance to redshift calibration which reduces the Poisson uncertainty on the transfer function for galaxies that are likely to be selected in the weak lensing (wide) sample. As the \Balrog transfer function already contributed negligible uncertainty to the final redshift estimate in DES Y3 \citep*{y3-wlpz}, we expect the Y6 uncertainty to also be negligible.

\subsection{Redshift calibration samples}
\label{sec:Redshift_calibration_samples}

We use a redshift calibration sample that overlaps the DES deep fields, primarily, the COSMOS field, in order to characterize the optical+NIR colour-redshift relation among galaxies that pass our weak lensing source sample selection in the wide field. Fig.~\ref{fig:deep field} highlights the redshift information that overlaps the DES optical+NIR deep data.

The strategy for building the redshift calibration sample is to maximize accuracy using the available high-resolution spectroscopy, while ensuring that the sample is complete and representative of the WL data by supplementing the spectra with narrow-band photometric redshifts. Spectroscopic data are highly incomplete at the faint magnitude limit of our source sample (24.7) and relying entirely on spectroscopic calibration would result in a biased $n(z)$ inference due to selection effects \citep[e.g.,][]{GruenBrimioulle2017}. We use spectra where available, followed by \texttt{PAUS} and then \texttt{COSMOS2020} elsewhere. 

Fig.~\ref{fig:zsamples hist} shows the DES wide field $i$-band magnitude distribution for each redshift bin for galaxies which have deep $ugriz\rm{JHK_s}$ photometry and redshift information. Each galaxy has been weighted by the \mdet response statistical shear weight, and a galaxy detection probability from \Balrog. While the spectroscopic compilation spans the largest area among the redshift catalogs, it is also the shallowest. The \texttt{COSMOS2020} catalog is the deepest, but also has the lowest redshift precision. Finally, the \texttt{PAUS} catalog is more precise than \texttt{COSMOS2020} and, unlike spectroscopic samples, is nearly complete in the highly relevant magnitude range of up to $i\approx23$ but has the lowest area coverage at faint magnitudes. As a result, for the higher redshift bins (right panels), which correspond to fainter galaxies, the calibration is dominated by \texttt{COSMOS2020}. The lack of spectroscopic redshift sample ($<20\%$ in the highest redshift bin in Fig.~\ref{fig:zsamples hist}) is a limiting factor in our redshift calibration and will worsen in deeper Stage IV surveys without improved data.

Here we briefly describe our choices in the spectroscopic, \texttt{PAUS}, and \texttt{COSMOS2020} catalog construction. Our spectroscopic catalog is built from C3R2 \citep{c3r2_2017,c3r2_2019,c3r2_2021}, zCOSMOS \citep{zcosmos}, VVDS \citep{vvds}, and VIPERS \citep{vipers}. PAUS is a 66-band photometric redshift catalog \citep{paus-cosmos} built from 40 narrow bands in PAU Survey data \citep{paus} and 26 COSMOS2015 bands excluding the mid-infrared \citep{cosmos2015}. There are two photometric catalogs in \texttt{COSMOS2020} \citep{cosmos2020}: \texttt{CLASSIC} -- the \texttt{COSMOS2015} aperture photometric
method, and \texttt{FARMER} -- the new profile-fitting photometric extraction method. Each have two redshift estimates: \texttt{LePhare} -- the  \texttt{COSMOS2015} single template fitting method, and \texttt{EAZY} -- a method with a different population of templates and using a non-negative linear combination of templates to fit the observed photometry. We use the \texttt{FARMER-LePhare} catalog, which outperforms other catalogs in the combination of precision and outlier fraction (see Figure 15 of \citealt{cosmos2020}), especially at $23<i<25$ where the vast majority of our galaxies reside for which \texttt{COSMOS2020} is providing the redshift information.

\subsection{DES Y6 image simulations}
\label{sec:DES_Y6_image_simulations}

A galaxy's flux is often "blended" with the flux from another galaxy that overlaps along the line of sight, particularly in deeper data. The blending of the images of galaxies of different redshifts results in a mixing of the shear responses and impacts the inferred shear and redshift calibration \citep{y3-blending, blending_li, blending_zhang}. To account for the effect, we rely on the image simulations described in \citet{y6-imagesims}. In brief, the wide-field images featuring galaxy and star populations with realistic source distributions are simulated for a series of redshift-dependent shear configurations. We use \texttt{COSMOS2015} \citep{cosmos2015} augmented by space-based (HST) imaging for bulge (de Vaucouleurs) + disk (exponential) model fits as our injection galaxy catalog, Gaia \citep{Gaia} as our stellar input, and the cell-based coadd method developed in \citet{Metadetect_lsst}.
We further simulate correct galaxy density and clustering based on the Cardinal simulations \citep{Cardinal}, as well as adding pixel-level systematics like \textsc{Piff} PSF models \citep{piff}, world coordinate solutions, and bright star masks derived from the real DES Y6 images.\footnote{The key difference between DES Y6 synthetic source sample and DES Y6 image simulations is the latter has fully simulated image properties that include the cell-based coadd, galaxy density and clustering correction, and pixel-level systematics.}

The DES Y6 image simulation suite includes ten redshift intervals: [0.0 - 0.3, 0.3 - 0.6, 0.6 - 0.9, 0.9 - 1.2, 1.2 - 1.5, 1.5 - 1.8, 1.8 - 2.1, 2.1 - 2.4, 2.4 - 2.7, 2.7 - 6.0]. For each simulation, a positive shear $g_1 = 0.02$ is applied to a redshift interval, while a negative shear $g_1 = -0.02$ is applied to the redshift range outside. There are also two simulations with constant $g_1 = 0.02/$-$0.02$ applied to all galaxies simultaneously. \mdet is then run (with identical configuration as for the DES Y6 data) on each of the sheared images to produce shape catalogs with an artificial shear signal. These simulations provide the inputs of the shear and redshift-dependent blending effect. The \mdet measured shear of these simulations on each redshift interval is compared with the true shear imposed. We found that blending affects the \mdet shear response, biasing galaxies in the sheared sample, and also induces a spurious positive response in neighboring galaxies, which leads to redshift-dependent blending effects across redshift intervals. This effect is modeled and corrected both in the weighted redshift distributions and in the shear multiplicative bias. More detail can be found in \citep{y6-imagesims} and in later sections.

\section{SOM-based redshift inference}\label{sec:SOMPZ}

This section outlines the primary method to estimate the DES Y6 redshift distributions: SOMPZ. We introduce the Self-Organizing Map (SOM) as an unsupervised machine learning algorithm and describe the training and assignment of the SOM using galaxy fluxes in Section \ref{sec:Methodology}. The DES Y6 redshift calibration scheme is built using an updated SOM approach that incorporates several improvements over the fiducial method adopted in Y3 \citep*{y3-wlpz}, which we describe in Section \ref{sec:SOM_Y6_improvements}. The SOMPZ methodology is validated using galaxy image simulations generated by \citet{y6-imagesims} in Appendix~\ref{app:imsim_nz_validation}.

\begin{figure*}
    \centering
    \includegraphics[width=\textwidth]{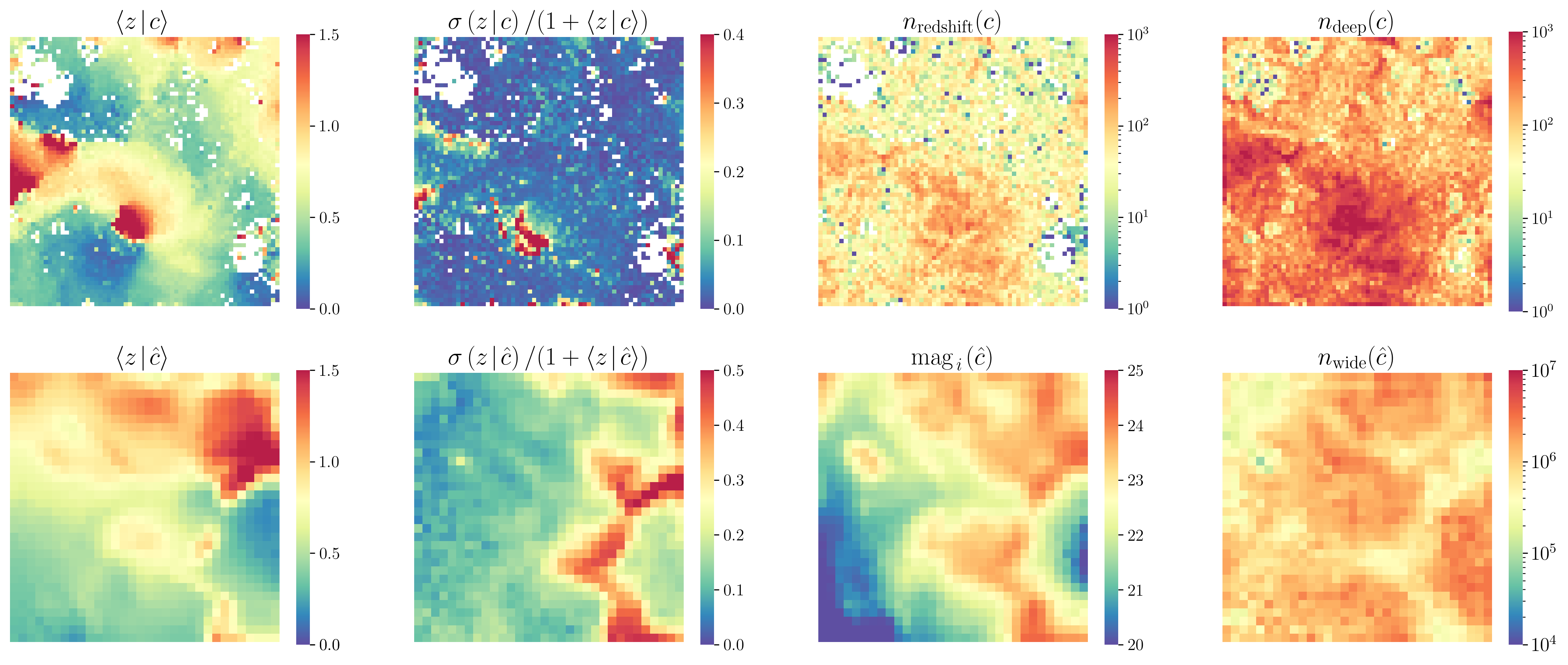}
    \caption{\label{fig:SOMs} The deep ($c$, top panels) and wide ($\hat{c}$, bottom panels) field Self-Organizing Maps. The SOMs are coloured to represent the mean redshift of each cell (left),  the standard deviation of the redshift distribution of each cell scaled by $(1+\langle z \rangle)$ (second), the weighted total number of calibration redshifts galaxies assigned to each SOM cell (third, top), the mean $i$-band magnitude (third, bottom) and the weighted total number of deep/wide DES galaxies assigned to each SOM (right). White cells in the deep SOM indicate parts of colour space for which there are no representative calibration sample galaxies, and the deep galaxies in those cells are discarded from the analysis.}
\end{figure*}

\subsection{SOMPZ Methodology} 
\label{sec:Methodology}

The SOM projects high-dimensional data -- in our case, the multi-band flux measurements of galaxies -- onto a two-dimensional grid of cells. After projection, each SOM cell can be interpreted as representing a phenotype of galaxies, that is, a group of galaxies with similar multi-band fluxes and therefore similar redshifts. In addition to grouping galaxies with similar fluxes into phenotypes, the SOM provides a topological visualization of transitions between different phenotypes across neighbor cells, thereby informing sample selection \citep{sompz_masters}. The SOMPZ process is summarized below and we refer the reader to \citet*{y3-wlpz} for more details.

(A) \underline{SOM training} to define the range of galaxy phenotypes: During training, the SOM is initialized by galaxies with multi-band flux measurements. Each SOM cell on the two-dimensional SOM grid also stores a representative set of fluxes, referred to as the cell flux. For each training galaxy, the algorithm (detailed in  Appendix~\ref{app:SOMF} and Section~\ref{sec:SOM_Y6_improvements}) finds the cell whose cell flux is most similar to the galaxy's flux. We call this the "winner" cell, which then updates its flux to better match the input galaxy flux, using a learning rate that decreases over training time. The cells that neighbor the winner cell also update their cell flux, weighted by a 2D Gaussian kernel neighborhood function centered on the winner. This results in a trained SOM where each cell represents a characteristic type of galaxy, while neighboring cells exhibit a smooth transition in flux. 

The "wide SOM" uses $32 \times 32$ cells and is built using the $griz$ fluxes of the DES Y6 weak lensing source sample. In order to break galaxy redshift-type degeneracies, we incorporate NIR information from the DES optical+NIR deep field data. Therefore, we use a second "deep SOM", which can be defined with higher resolution of $64 \times 64$ SOM cells given the 8-band $ugriz\rm{JHK_s}$ deep information. The deep and wide SOMs are trained using 3 million draws of the DES optical+NIR deep data and 10 million draws of the Y6 weak lensing source sample, respectively.

(B) \underline{SOM assignment} to identify which cell a \mbox{galaxy falls into}: Individual galaxies are assigned to the SOM cells whose cell flux most closely matches the galaxy flux. The optical+NIR deep data are assigned to the deep SOM, and we estimate the cell probability $p(c)$ as the fraction of galaxies falling into the respective cell $c$. Similarly, the source sample is assigned to the wide SOM cells $\hat{c}$ and we estimate $p(\hat{c})$ proportional to the number of galaxies in that sample falling into the respective cell.

(C) \underline{Redshift mapping}: The redshift sample is \mbox{assigned to the deep} SOM to define the redshift probability given a deep SOM cell, $p(z|c)$. The SOMPZ redshift calibration is a reweighting process. The redshift probability distribution for an ensemble of galaxies can be obtained by a sum over deep cells, $c$:
\begin{equation}
\label{eqn:pzdeep}
p(z_\text{deep}) = \sum_{c} p(z|c) p(c) \, ,
\end{equation}
where the same selection procedure is assumed to be applied to the deep galaxies and the ensemble of galaxies in question. The first reweighting uses the deep sample with $ugriz\mathrm{JHK_s}$ bands to break colour-redshift degeneracy.

(D) \underline{Transfer function}: The synthetic source sample is assigned to the deep and wide SOM to define the probability $p(c|\hat{c})$ of the galaxies in wide SOM cell $\hat{c}$ to reside in deep SOM cell $c$. The second reweighting uses this transfer function $p(c|\hat{c})$ to connect the deep and wide SOM, and thus to infer the probability distribution of weak lensing source galaxies in the wide SOM, following
\begin{equation}
\label{eqn:pzwide}
p(z_{\rm wide}) = \sum_{\hat{c}} \sum_{c}p(z|c,\hat{c}) p(c|\hat{c}) p(\hat{c}).
\end{equation}

In Fig.~\ref{fig:SOMs} we show the fiducial DES Y6 deep and wide SOMs, coloured to represent various properties. In the wide SOM, galaxy properties vary smoothly across neighboring cells and the galaxies are distributed evenly. The deep SOM, which has additional bands to break degeneracies, shows a few sharp features where transitions between types happen,  and thus galaxy density is less evenly distributed. In the second-column scatter plots, our deep SOM cells have a low redshift scatter, demonstrating the deep SOM's role in breaking colour-redshift degeneracy, with a larger scatter at higher redshift and fainter magnitudes.

(E) \underline{Redshift bins:} We divide the weak lensing source galaxies into four redshift bins $b$, containing roughly equal number of galaxies. In particular, redshift bin edges are defined as $[0.0, 0.40,0.68,0.96,3.0]$, and each wide SOM cell is assigned to the bin according to the mode of its redshift distribution. The redshift distribution of a bin is then defined as: 
\begin{equation}
\label{eqn:pz_full}
p(z | b) = \sum_{\hat{c} \in b} \sum_{c}  \underbrace{p(z | c, \hat{c})}_{\text{Redshift}} 
\underbrace{p(c | \hat{c})}_{\text{Deep-Balrog}} 
\underbrace{p(\hat{c} | b)}_{\text{Wide}}.
\end{equation}
In \citet{Roster_euclid}, it was found that defining tomographic bins using the mean photometric redshift of individual sources provides greater fidelity in isolating galaxies that truly belong to the desired tomographic redshift range, compared to a purely SOM-based approach that relies on the mean redshift of individual SOM cells. However, this result strongly relies on the availability of unbiased photometric redshift estimates -- a condition we cannot realistically assume. Furthermore, \citet{kang_prep} showed that the most important step to achieve a precise and unbiased SOM calibration of the bins is to include a realistic transfer function (here \Balrog), a step that was not included in \citet{Roster_euclid}. They also demonstrated that once this is accounted for, the improvement reported by \citet{Roster_euclid} becomes marginal for realistic DES-like photometry, corresponding to Euclid DR1. 

Given the large number of $c$, $\hat{c}$ pairs, there are often few to no redshift galaxies that satisfy both conditions in $p(z | c, \hat{c})$. We make the assumption that the redshift of galaxies inside a deep cell $c$ shall not be highly sensitive to its noisy wide field photometry. This way, we relax the condition on $\hat{c}$ to $b$, dubbed "bin-conditionalization":
\begin{equation}
\label{eqn:pz_bc}
p(z|b) \approx \sum_{c, \hat{c}} \underbrace{p(z|c,b)}_{\text{Redshift}}  \underbrace{p(c|\hat{c})}_{\text{Deep-Balrog}} \underbrace{p(\hat{c}|b)}_{\text{Wide}}.
\end{equation}
In instances where no redshift galaxy satisfies the conditions in $p(z|c,b)$, we relax further to $p(z|c)$.

Each weak lensing source galaxy is weighted by its statistical shear weight and response weight $ w = w_{\text{stat}}  R$ (defined in Section~\ref{sec:DES_Y6_weak_lensing_source_sample}). We also weight each \Balrog galaxy by the inverse of its injection count in the simulation: $ w = w_{\text{stat}}  R/ N_{\rm inj}$. Each redshift or deep galaxy is injected multiple times into the wide field, and we treat each detection $j$ with its \Balrog weight
$w_j = w_{\text{stat},j} R_j / N_{\rm inj}$. 
The total weight of the galaxy in tomographic redshift bin $b$ is therefore the sum over its detections in that bin
$ w = \frac{1}{N_{\rm inj}} \sum_{j=1}^{N_{\rm det}(b)} w_{\text{stat},j} \, R_j .$

\subsection{Y6 advancements}

\label{sec:SOM_Y6_improvements}

Unlike the Y3 analysis, the DES Y6 redshift calibration scheme is built on the SOMF (Self-Organizing Map for faint galaxies) method described in \citet*{svsn} and \citet{somf}. This method is designed to better characterize the faint galaxies in our deeper lensed source sample and to consider imperfect training data. Here we summarize the two most important improvements, and we give more detail in Appendix~\ref{app:SOMF}.

First, SOMF considers the signal-to-noise, $S/N$, of each galaxy's band during the training phase, penalizing low $S/N$ photometry to ensure that galaxies with higher $S/N$ in a given band have more influence on the metric. This reduces the impact of noisy measurements, enabling a more robust characterization of galaxy phenotypes. Second, instead of using lupticolours and luptitudes, as in the Y3 analysis, SOMF uses galaxies' fluxes to group galaxies into SOM cells. While colour is in general more redshift sensitive, at low $S/N$ colour measurements are dominated by noise, leading to large uncertainties. In addition, C3R2 \citep{c3r2_2017, c3r2_2019, c3r2_2021, dc3r2} have found that for galaxies with a fixed colour, there is still a dependence of redshift on magnitude. Flux, on the other hand, is more robust on "noisy" galaxies with low $S/N$ ratios or even negative flux values. Together, it was demonstrated in \citet{somf} that SOMF reduces the overlap in the resulting redshift distributions.

\begin{figure}
    \centering
    \includegraphics[width=\columnwidth]{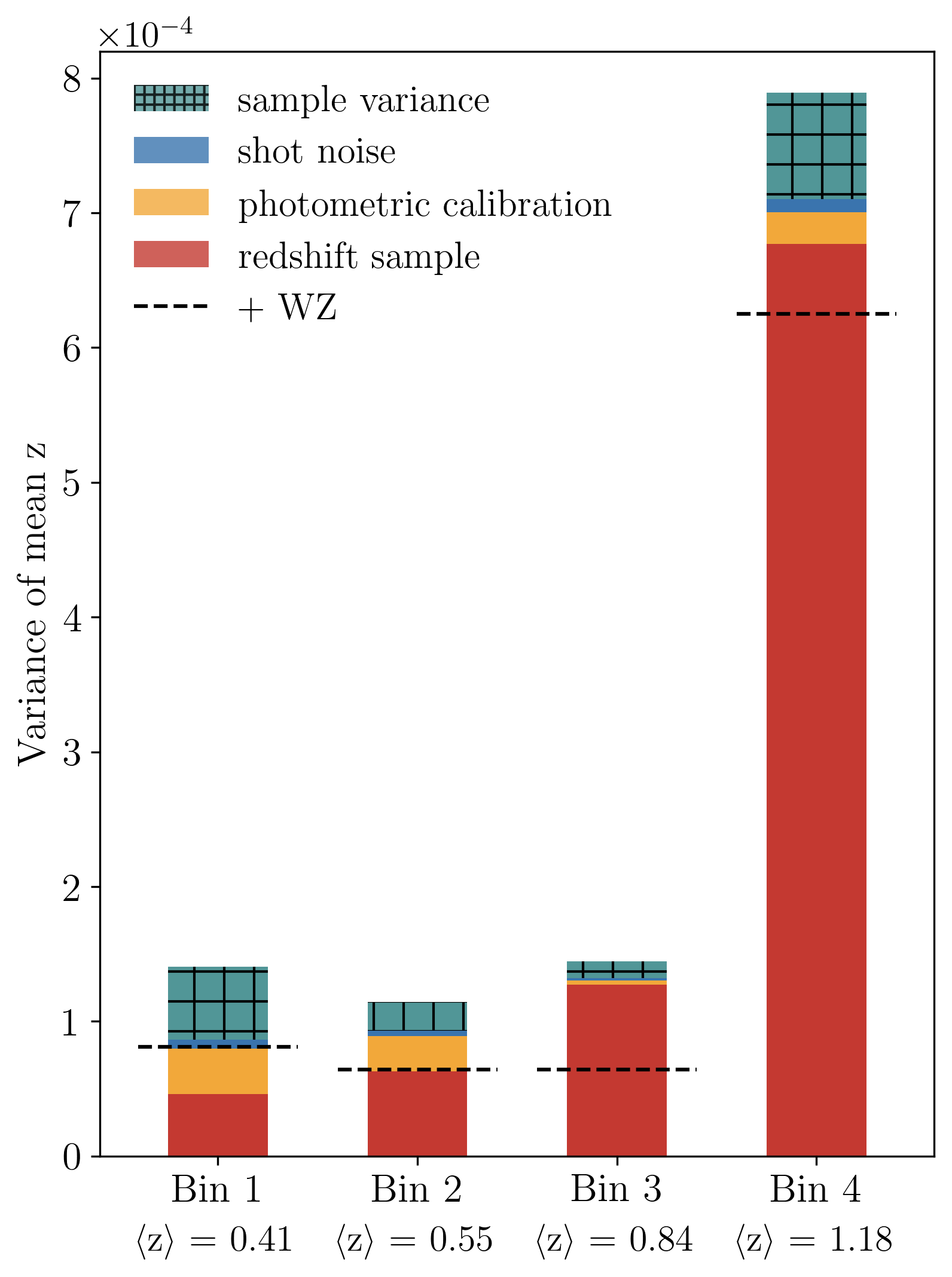}
    \caption{\label{fig:uncertainty_bar} The relative contribution of each source of uncertainty for the four redshift bins. The variance due to the calibration redshift sample (red) is dominant, followed by the sample variance (green). Compared to these two, the variance due to photometric calibration (yellow) and shot noise (blue) are small. The dashed black lines show the constraints on mean redshift after adding clustering information. The variance of mean redshift is shown in the plot to illustrate the relative magnitude of each source of uncertainty in each bin and their evolution with redshift; the corresponding uncertainties are reported in Table~\ref{tab:uncertainty}.} 
    
\end{figure}

\begin{table*}
\centering
\begin{tabular}{l@{\hskip 2em}cc@{\hskip 2em}cc@{\hskip 2em}cc@{\hskip 2em}cc}
                    & \multicolumn{2}{c}{Bin 0 \makebox[1.2em][r]} & \multicolumn{2}{c}{Bin 1 \makebox[1.2em][r]} & \multicolumn{2}{c}{Bin 2 \makebox[1.2em][r]} & \multicolumn{2}{c}{Bin 3} \\
                    & $\langle z \rangle$ & $\sigma_{\langle z \rangle}$ 
                    & $\langle z \rangle$ & $\sigma_{\langle z \rangle}$ 
                    & $\langle z \rangle$ & $\sigma_{\langle z \rangle}$ 
                    & $\langle z \rangle$ & $\sigma_{\langle z \rangle}$ \\
\hline
Y6 SN               & 0.420 & 0.003 & 0.553 & 0.002 & 0.843 & 0.001 & 1.182 & 0.003 \\
Y6 SVSN             & 0.416 & 0.008 & 0.549 & 0.005 & 0.841 & 0.004 & 1.178 & 0.009 \\
Y6 SVSN + ZPU       & 0.417 & 0.010 & 0.549 & 0.008 & 0.841 & 0.006 & 1.179 & 0.010 \\
Y6 SVSN + ZPU + RU  & 0.415 & 0.012 & 0.548 & 0.011 & 0.842 & 0.012 & 1.179 & 0.028 \\
\hline
Y6 SOMPZ            & 0.415 & 0.012 & 0.548 & 0.011 & 0.842 & 0.012 & 1.179 & 0.028 \\
Y6 SOMPZ + WZ       & 0.428 & 0.009 & 0.561 & 0.008 & 0.849 & 0.008 & 1.191 & 0.025 \\
Y6 SOMPZ + WZ + blending & 0.415 & 0.012 & 0.540 & 0.009 & 0.847 & 0.009 & 1.157 & 0.025 \\
Y6 SOMPZ + WZ + blending + mode & 0.414 & 0.012 & 0.538 & 0.008 & 0.846 & 0.009 & 1.157 & 0.024 
\end{tabular}

\caption{Values of and approximate error contributions to the mean redshift of each redshift bin for each step in the methodology. We find that the error due to the uncertainty in the redshift calibration sample is the dominant source of uncertainty, followed by the sample variance in the deep fields. On the other hand, WZ uses additional clustering information to better constrain mean redshift uncertainty.
}
\label{tab:uncertainty}
\end{table*}

\section{Characterizing Redshift Uncertainties}
\label{sec:redshift_calibration_uncertainty}
In this section we summarize the methodology used to characterize sources of uncertainty in the DES Y6 redshift calibration, and present their impact on the SOMPZ $n(z)$. In the following subsections, we describe in more detail the model and implementation to characterize each uncertainty and discuss their relative contributions.

In summary, we generate $10^8$ realizations of the lensed source sample's $n(z)$ that span the possible instances of redshift distributions we might have estimated in a separate realization of the experiment, which reflects the following sources of uncertainty. 

(1) \underline{Redshift calibration sample bias (RU)} due to \mbox{imperfect} photometric calibration data from \texttt{COSMOS2020} and \texttt{PAUS}, particularly for faint, high-redshift galaxies. We model these via coherent shifts in magnitude-redshift bins based on spectroscopic comparisons. We draw a single Latin Hypercube (\texttt{LHC}) sample for each of the two redshift samples that shifts the \texttt{COSMOS2020} and \texttt{PAUS} redshifts to account for redshift sample bias (Section \ref{sec:redshift_sample_bias}).

(2) \underline{Photometric zero-point calibration uncertainty (ZPU)} in the deep fluxes due to field-to-field variations. We model these by perturbing the deep photometry within the SOMPZ framework. A single LHS is drawn for each of the 24 deep field inputs (3 fields × 8 fluxes) that shifts the deep flux to account for the calibration uncertainty (Section \ref{sec:deep_field_photometric_calibration_uncertainty}). 

Sources (1) and (2) are propagated at the same time to compute a single SOMPZ $n(z)$ realization for the source galaxies. In total, 100 realizations are generated.

(3) \underline{Sample variance (SV)}, which arises from limited area \mbox{of the} redshift and deep field observations that do not fully represent the fluctuations in the matter density field (Section \ref{sec:sv_and_sn}). 

(4) \underline{Shot noise (SN)}, which captures the Poisson \mbox{fluctuation due} to limited number of galaxies (Section \ref{sec:sv_and_sn}). 

Sources (3) and (4) are jointly propagated using the three-step Dirichlet (3sDir) analytical model. For each of the 100 $n(z)$ realizations from (1) and (2), we further generate 1,000,000 realizations using 3sDir to account for sample variance and shot noise (Section \ref{sec:sv_and_sn}). 

(5) \underline{Blending (BL)} of nearby galaxies, which introduces a redshift-dependent bias in measuring galaxy response. We measure this using image simulations, and model uncertainties due to simulation imperfections. This bias and uncertainty is applied after further constraining SOMPZ $n(z)$s with clustering information, described in Section \ref{sec:blending}. 

Compared to the Y3 analysis of \citet*{y3-wlpz}, we do not include uncertainty from the SOMPZ methodology, as it is shown to be negligible in Appendix~\ref{app:imsim_nz_validation}. We also find that with the enhanced injection scheme and a total of 10,000 tiles, the uncertainty due to the \Balrog process is negligible in Y6.

The SOMPZ result for each redshift bin from the 100,000,000 $n(z)$ realizations that span the main sources of uncertainty (except for blending), is shown in Fig.~\ref{fig:violins2} as black violins. We illustrate the relative contributions for each redshift uncertainty component across the four redshift bins as bar plots in Fig.~\ref{fig:uncertainty_bar} and numerically in Table~\ref{tab:uncertainty}.

\subsection{Redshift Calibration Sample Imperfections}
\label{sec:redshift_sample_bias}

\begin{figure*}
    \centering
    \includegraphics[width=\textwidth]{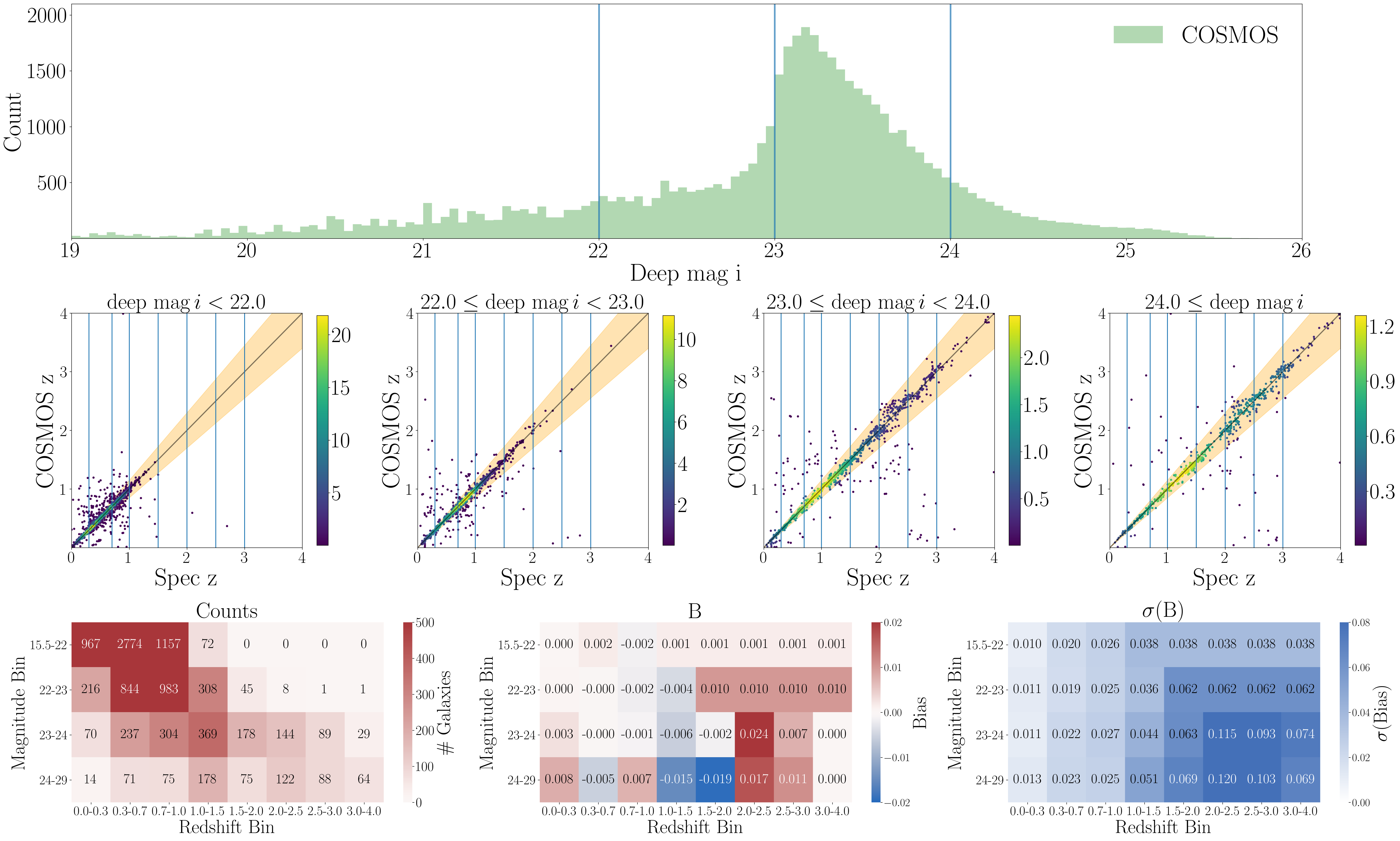}
    \caption{\label{fig:redshift_sample_uncertainty_hist} Illustration of the construction of the redshift calibration sample uncertainty. We model the uncertainty due to imperfect photometric redshifts from \texttt{COSMOS2020} as a bias function of $i$-band magnitude and redshift, based on the direct comparison with spectroscopic redshifts, where available. The upper panel shows the weighted distribution of the DES deep field $i$-band magnitude for the lensing selection that is calibrated by \texttt{COSMOS2020}, showing that this calibration sample is most important for the two upper magnitude bins. We inspect the agreement of spectroscopic and \texttt{COSMOS2020} photometric redshifts in four magnitude bins (corresponding to each column, middle panel) and eight redshift bins. In the lower panel, we report the number of galaxies with both spectroscopic redshift and \texttt{COSMOS2020} photometric redshift information (left), the median redshift bias of the \texttt{COSMOS2020} photometric redshift with respect its spectroscopic match (middle) and the uncertainty on redshift bias (right). For bins containing fewer than 10 galaxies, the median and uncertainty of redshift bias are assigned to the values from the adjacent lower redshift bin. 
    }
\end{figure*}

In the ideal situation, the redshift calibration sample would consist of only spectroscopic galaxies. However, spectroscopic data is incomplete, especially at higher redshift and fainter magnitudes. To avoid the selection biases incurred from a spectroscopic-only calibration sample, it is necessary to include photometric redshifts from narrow-band data, such as \texttt{COSMOS2020} and \texttt{PAUS}. The upper panel of Fig.~\ref{fig:redshift_sample_uncertainty_hist} shows the $i$-band magnitude distribution for the subset of the optical+NIR deep data that is calibrated by \texttt{COSMOS2020}. Despite having high completeness, this photometric sample, which peaks at $i$-magnitude$>$22.5, has photometric redshifts that are not as accurate as their spectroscopic counterparts. This is the sample (and equivalent for \texttt{PAUS}) for which we compute and assign a redshift sample uncertainty in our analysis. 

\noindent{\underline{\textbf{Modeling}}}: We model the redshift bias of the photometric sample \texttt{COSMOS2020} and \texttt{PAUS} by considering their subsample that also has a spectroscopic redshift. For clarity, we refer to the subset of \texttt{COSMOS2020} that is also observed in the spectroscopic sample as the \texttt{spec-COSMOS} sample and that for \texttt{PAUS} as \texttt{spec-PAUS}. 

We divide the \texttt{spec-COSMOS} sample into four subsamples by $i$-band magnitude, and then further into eight \texttt{COSMOS} redshift bins, indicated in the middle row of Fig.~\ref{fig:redshift_sample_uncertainty_hist}. These middle four panels present the spectroscopic, $z_\mathrm{spec}$, versus photometric redshift, $z_\mathrm{photo}$, for \texttt{spec-COSMOS}. 
By computing the difference between each galaxy’s photometric redshift and spectroscopic redshifts in these bins, we can model the bias in the \texttt{spec-COSMOS} sample as a function of magnitude and redshift. 
For each magnitude-redshift bin, the bias, $B$, is defined as
\begin{equation}
B = \textrm{median}(z_{\rm spec} - z_{\rm photo}).
\end{equation}
Because spec-COSMOS is not fully representative of COSMOS at the same magnitude, the bias measured in spec-COSMOS may not be the same as that in COSMOS. As a partial mitigation of this, we therefore define an uncertainty on the bias as
\begin{equation}
\sigma_{\rm B}=\textrm{std}(z_{\rm spec} - z_{\rm photo}) .
\end{equation}
Following the criteria of \citep{cosmos2020}, we define outlier galaxies as those that satisfy the condition $(z_{\rm spec} - z_{\rm photo})/z_{\rm spec}>0.15$, indicated as outside of the yellow shaded region, and exclude these from this estimation. The bottom row quantifies the count of \texttt{spec-COSMOS} in each magnitude (row) and redshift (column) bin, as well as the median bias, $B$ and the standard deviation, $\sigma_B$. An equivalent procedure is performed for the \texttt{spec-PAUS} sample, except with different bin edge choices as \texttt{PAUS} contains fewer galaxies and is generally brighter and at lower redshifts compared to \texttt{COSMOS2020}. We use these statistics as a model of the redshift bias of the \texttt{COSMOS2020} and \texttt{PAUS} samples. 

\noindent{\underline{\textbf{Propagation}}}:
For each galaxy in \texttt{COSMOS2020} (and equivalently for \texttt{PAUS}), we assign the appropriate bias according to its $i$-magnitude and redshift. To incorporate this uncertainty in the computation of the lensed source sample's $n(z)$, within the SOMPZ scheme, we apply the redshift sample bias by perturbing the \texttt{COSMOS2020}'s redshifts,  $z_{\texttt{COSMOS2020}}$, by a Gaussian draw, assigning each galaxy's redshift to be
\begin{equation}
z = z_{\texttt{COSMOS2020}} + G(B,\sigma_{\rm B}) \, ,
\end{equation}
where the Gaussian error is generated using Latin Hypercube sampling (\texttt{LHS}) among all individual galaxies. LHS is a sampling method that stratifies each parameter dimension into equally probable intervals and reduces estimator variance compared to simple random sampling \citep{LHC}. In other words, the same number of LHS points maps out the parameter space more accurately than an equal-sized random sample. Note that the same LHS point is used for all galaxies in \texttt{COSMOS2020} to enable a coherent shift of bias across the sample. We generate a total of 200 LHS points (100 for \texttt{COSMOS2020} and 100 for \texttt{PAUS}) to provide a well-sampled full Gaussian distribution. The resulting 100 $n(z)$ realizations encapsulate the possible redshift sample biases.

\noindent{\underline{\textbf{Results}}}: This uncertainty, shown in red in Fig.~\ref{fig:uncertainty_bar}, is the dominant source of uncertainty in the DES Y6 redshift calibration, which makes sense given the substantially deeper source sample. Compared to Y3, our $i$-band limiting magnitude increases from 23.5 to 24.7. The redshift sample uncertainty increases with redshift. At higher redshifts, we rely more on photometric redshift sample galaxies from \texttt{COSMOS2020} and \texttt{PAUS} as spectra make up only 19\% of the calibration sample, compared to 48\% in the first bin (Fig.~\ref{fig:zsamples hist}). Compounding this, the quality of \texttt{COSMOS2020} and \texttt{PAUS} further decrease at higher redshifts and fainter magnitudes, such that the uncertainty on bias compared to spectroscopic counterparts increases (Fig.~\ref{fig:redshift_sample_uncertainty_hist}).

\subsection{Deep Field Photometric Calibration Uncertainty}
\label{sec:deep_field_photometric_calibration_uncertainty}

The photometric calibration of the deep field data varies across the four fields. This is encoded as zero point uncertainty for each band. We are using the same deep field and photometric calibration uncertainty methodology as \citet*{y3-wlpz}. 

\noindent{\underline{\textbf{Modeling}}}: The uncertainty for $ugriz\mathrm{JHK_s}$ bands in  magnitudes are: $\sigma_\mathrm{ZP}=$ [0.055, 0.005, 0.005, 0.005, 0.005, 0.008, 0.008, 0.008], respectively \citep{y3-deep}. We include these uncertainties and note that the $u$-band uncertainty is an order of magnitude larger than any other bands.

\noindent{\underline{\textbf{Propagation}}}: For each realization, we generate a zero-point bias as a Gaussian draw centered on zero with a width that reflects the uncertainty of each band, $G(0,\sigma_\mathrm{ZP})$, using the LHS method. As only the relative difference in photometric zero point matters in uncertainty propagation, we fix the deep fluxes in the COSMOS field and perturb those in the X3, C3, and E2 fields, generating three sets of LHS samples for 8 bands. The $n(z)$ is calculated using the SOMPZ method with the perturbed deep field galaxy fluxes, and this process is repeated to produce 100 $n(z)$ realizations.

\noindent{\underline{\textbf{Results}}}: We find that this deep field photometric calibration uncertainty (yellow in Fig.~\ref{fig:uncertainty_bar}) is largest in Bin 1 because the $u$-band, where the calibration uncertainty is most significant, has the strongest influence on low redshift galaxies. The uncertainty decreases in Bin 2 and 3 but increases again in Bin 4. Because redshift is approximately proportional to the exponential of apparent magnitude, the same uncertainty in magnitude leads to a larger redshift uncertainty at higher redshift.

\subsection{Sample Variance \& Shot Noise}
\label{sec:sv_and_sn}
Shot noise arises from the finite number of galaxies we observe in redshift and deep fields, introducing Poisson noise on the number of observed galaxies for any combination of colour and redshift. On the other hand, sample variance arises from the limited area of the redshift and deep field, due to fluctuations of the underlying matter density. We use the same sample variance and shot noise calibration methodology as 
\citet*{y3-wlpz} and \citet*{svsn}.

\noindent{\underline{\textbf{Modeling}}}: Under the influence of sample variance and shot noise, the number of redshift sample galaxies in redshift bin $z$ is modeled to be
\begin{equation}
N_z = \text{Poisson} \left[ N f_z (1 + \delta_z) \right],
\end{equation}
where $N$ is the total number of galaxies in the redshift sample, and $f_z$ is the normalized redshift distribution of the redshift sample. The Poisson distribution is used for modeling shot noise, while $\delta_z$ captures the variation in galaxy over-density at $z$ due to sample variance. From this equation, it is straightforward to derive using the law of total variance
\begin{equation}
\begin{aligned}
\mathrm{Var}(N_z) 
&= \mathrm{E}\big[\mathrm{Var}(N_z\mid \delta_z)\big] 
  + \mathrm{Var}\big(\mathrm{E}\,[N_z \mid \delta_z]\big) \\
  &=\mathrm{E}[N f_z(1 + \delta_z)] + \mathrm{Var}(N f_z(1 + \delta_z)) \\
  &=N f_z + N^2 f_z^2 \mathrm{Var}(\delta_z)
\end{aligned}
\end{equation}
\vspace{-1.2em}
\begin{equation}
\langle N_z \rangle = N f_z,
\end{equation}
which is then used to derive the normalized variance for each redshift bin z,
\begin{equation}
\lambda_z \equiv \frac{\text{Var}(N_z)}{\langle N_z \rangle} = 1 + N f_z \text{Var}(\delta_z).
\end{equation}
\
Here $\langle N_z \rangle$ is the variance in shot noise, and acts as the normalizing factor. The unity term represents the normalized shot noise and the term $Nf_z \text{Var}(\delta_z)$ represents the normalized sample variance. This normalized variance will be very useful when we propagate sample variance and shot noise in the source galaxy redshift calibration.

We use a theoretical method to calculate $\text{Var}(\delta_z)$: 
\begin{equation}
\text{Var}(\delta_z) \equiv  \sum_l \frac{2l+1}{4\pi}F^2_lC^{zz}_l.
\end{equation}
$\text{Var}(\delta_z)$ is independent of calibration sample galaxy density. It is calculated by first inferring the matter power spectrum using \texttt{CAMB}. Then we use \texttt{COSMOS2015} data to correct for the galaxy-matter bias by computing the galaxy correlation function within the small patch and comparing it with the theory prediction.
Combining the matter power spectrum with the galaxy-matter bias yields the galaxy power spectrum $C_l$. The galaxy power spectrum is further smoothed at scales of the redshift or deep field area, keeping only the large scale modes beyond the scale of the deep field, by using the top hat smoothing function $F_l$. The smoothed galaxy power spectrum is summed up and this theoretically calculated sample variance is validated using \texttt{MICE} simulations \citep*{svsn}. In this way, we can calculate two $\text{Var}(\delta_z)$ with different field area: redshift field $\text{Var}^R(\delta_z)$ and deep field $\text{Var}^D(\delta_z)$. Since the redshift and deep fields in DES Y6 cover the same area as those in DES Y3, we adopt them from \citet*{y3-wlpz}. We get
\begin{equation}
    \begin{aligned}
    &\lambda^R_z \equiv  1 + N^R_z \text{Var}^R(\delta_z) \\
    &\lambda^D_z \equiv  1 + N^D_z \text{Var}^D(\delta_z)
    \end{aligned}
\end{equation}
for redshift and deep field, where $N^D_z$ is calculated by reweighting $N^R_z$ according to the deep to redshift galaxy occupation ratio within each deep SOM cell. The two quantities will be key components in uncertainty propagation in the next section.

\noindent{\underline{\textbf{Propagation}}}:
We use Dirichlet sampling -- a sampling method that preserves the total sample size in normalized histograms -- to propagate sample variance and shot noise. This is ideal in our case, since we want the galaxies in each deep SOM cell and redshift interval to vary in response to sample variance and shot noise, but the total number of redshift and deep galaxies need to stay the same. 

As an illustration of use, in a deep cell c we have $\vec{N} = [N_{1}, N_{2}, ..., N_{{max}}]$, where $N_{i}$ is the number of redshift sample galaxies in redshift bin $z_i$ of cell $c$. We then want to generate possible realizations of $\vec{N}$ taking into account Poisson distributed shot noise:

\begin{equation}
\vec{p} \leftarrow \mathrm{Dir}(\vec{N}) \propto \prod_{i=1}^{{\rm max}} p_{i}^{N_{i}-1}.
\end{equation}
We can draw the probability $\vec{p}$ from this Dirichlet distribution, where each $p_{i}$ is the probability that the redshift sample galaxy falls in redshift bin i. By virtue of the Dirichlet sampling, each draw of $\vec{p}$ satisfies i) $p_i>0$ and ii) $\sum_i p_i = 1$. A realization of $\vec{N}$ is obtained by multiplying {$\vec{p}$} by the total number of redshift sample galaxies $N_{\rm tot}$ in cell c.

In this simplified case, we can get a glimpse of how we propagate sample variance into additional shot noise. There are two properties of the Dirichlet distribution:
\begin{equation}
    \begin{aligned}
    &\langle p_i \rangle = N_i/N_{\rm tot} \\
    &\text{Var}(p_i) = N_i/N^2_{\rm tot}
    \end{aligned} \; .
\end{equation}
By rescaling $\vec{N}$, i.e.~changing all $N_i$ and $N_{\rm tot}$ by the same factor, we can modulate our confidence level on $\vec{p}$: when $N_i \rightarrow N_i/ \lambda$, then $\rm{Var}(p_i) \rightarrow \lambda \rm{Var}(p_i)$. If no rescaling is done ($\lambda = 1$), we are including only shot noise in our sampled $\vec{p}$, and the Dirichlet sampling is Poisson like. Taking $\lambda$ to be the ratio of sample variance and shot noise, which is closely related to the $\lambda_z^R$ and $\lambda_z^D$ we define above, we can include sample variance.

Following equation~(\ref{eqn:pz_bc}), we can expand: \footnote{$p^D(c)$ is the same as $p^B(c)$, since the \Balrog sample is the detection of injected deep galaxies in the wide field.}
\begin{equation}
p(z|b) \approx \sum_{\hat{c} \in b} \sum_c \frac{p^R(z,c)}{p^R(c)}p^D(c)\frac{p^B(c, \hat{c})}{p^B(c) p^B(\hat{c})}p^W(\hat{c}).
\end{equation}
where $c$ denotes the deep SOM cell and $\hat{c}$ denotes the wide SOM cell. The labeling $\{R, D, B, W\}$ represents the redshift sample, deep sample, \Balrog sample, and wide (source) sample, respectively.

There is one more step before we Dirichlet sample the redshift and deep sample probabilities. The number of galaxies found in adjacent deep cells $c$, or phenotypes, are correlated. Imagine in the limited area of the deep fields there is an excess of galaxies at a redshift $z$. A deep cell $c$ with $p(z)$ peaked at this redshift will have an excess number of galaxies. The neighboring deep cells of $c$, which have similar colours and a noticeable redshift overlap with $c$, will also likely be populated by a higher than usual number of galaxies. Thus, when we analyze sample variance, we cannot treat each phenotype as uncorrelated. Following Y3, this correlation is treated by grouping deep cells with similar mean redshift together to form superphenotypes. In this way, in one Dirichlet sampling instance where superphenotype $T$ fluctuates to have a larger number of galaxies due to sample variance, all phenotypes or deep cells $c$ within $T$ will coherently experience this positive shift in galaxy count. The equation can be further expanded with superphenotype $T$ as
\begin{equation}
\label{eq:3sDir_probexpand}
\begin{aligned}
p(z|b) =
&\biggl[\sum_{\hat{c} \in b} \sum_c \left( \sum_T p^R(c|z,T)p^R(z|T)p^R(T) \right) \frac{1}{p^R(c)} \\
& \cdot  \left( \sum_Tp^D(c|T)p^D(T)\right)
\frac{p^B(c, \hat{c})}{p^B(c) p^B(\hat{c})}p^W(\hat{c}) \biggr]. 
\end{aligned}
\end{equation}
The sum over $T$ is written explicitly here because in the next step all $T$s are sampled together from the Dirichlet distribution.

In addition to the correlation between adjacent deep cells, nearby redshift bins are also correlated. A limitation of the Dirichlet sampling method is that it assumes no correlation. To mitigate this, we use broader redshift bins of width 0.05 (see Figure 14 in \citet*{svsn}). Thus, for the calibration of sample variance and shot noise, the first bin of our redshift distribution is set to $[0.01,0.06]$. The start at $z = 0.01$ is because when $z\to 0$, sample variance diverges, while in practice this divergence is negligible since $p(z)\to 0$ at $z\to 0$.

We now sample the redshift and deep field probabilities using the Dirichlet distribution. This yields an $n(z)$ we might have measured in another realization of sample variance and shot noise. For each of the 100 calibration catalogs perturbed by redshift calibration sample imperfection and deep field photometric calibration uncertainty, we repeat the Dirichlet sampling process $10^6$ times to get a total of $10^8$ $n(z)$ realizations using
\begin{equation}
\begin{aligned}
p(z|b) \leftarrow
& \biggl[\sum_{\hat{c} \in b} \sum_c \left( \sum_T \text{Dir}\left(N^R_{zcT}\right)\text{Dir}\left(\frac{N^R_{zT}}{\lambda^R_T}\right)\text{Dir}\left(\frac{N^R_{T}}{\bar{\lambda}^R}\right) \right) \frac{1}{p^R(c)} \\
& \cdot \left( \sum_T \text{Dir}\left(N^D_{cT}\right)\text{Dir}\left(\frac{N^D_{T}}{\bar{\lambda}^D}\right) \right) \frac{p^B(c, \hat{c})}{p^B(c) p^B(\hat{c})}p^W(\hat{c})\biggr].
\end{aligned}
\end{equation}
Here $N$ is the number count of galaxies, where $N^R_{zcT}$ for example represents the number of redshift sample galaxies in redshift bin $z$, deep SOM cell $c$, and superphenotype $T$. Each Dirichlet sampled probability is further weighted by the bin-conditionalization proportionality and response+statistical weight to match equation~(\ref{eq:3sDir_probexpand}). The lambdas are calculated using the previously defined $\lambda_z^R$ and $\lambda_z^D$:
\begin{equation}
    \begin{aligned}
    &\lambda^R_T \equiv  \sum_{z} \lambda_z^R \frac{N_{zT}^{\mathcal{R}}}{N_T^{\mathcal{R}}} \\
    &\bar{\lambda}^R  \equiv \sum_{z} \lambda_z^R \frac{N_z^{\mathcal{R}}}{N^{\mathcal{R}}} \\
    &\bar{\lambda}^D \equiv \sum_z\lambda^D_z \frac{N^R_z}{N^R} 
    \end{aligned}
\end{equation}

\noindent{\underline{\textbf{Results}}}: We find that sample variance (green in Fig.~\ref{fig:uncertainty_bar}) is large for Bin 1 because, at low redshift, the limited area of the deep field ($\sim\! 10 \,\text{deg}^2$) and redshift field ($\sim \!1.38 \,\text{deg}^2$) captures less large-scale structure. The sample variance decreases in Bin 2 and Bin 3 but increases again in Bin 4. This arises from the tomographic binning scheme. We constructed our tomographic bins to have similar total galaxy counts. However, due to colour-redshift degeneracy, we could not cleanly separate galaxies by redshift. This causes the redshift distributions of the tomographic bins to be broad, especially in Bin 4, which contains mostly faint, high-redshift galaxies. When the redshift distribution spans a broader range, fluctuations farther from the mean redshift have a greater impact on the mean redshift, leading to a larger uncertainty. 

The shot noise (blue in Fig.~\ref{fig:uncertainty_bar}) is relatively small in all four tomographic bins. It is mainly dependent on the number of galaxies in each $z$ bin. Thus, in the tomographic bins with broader redshift distributions, the number of galaxies in each $z$ bin is lower, leading to larger shot noise, while fluctuations farther from the mean redshift further amplify this effect.

\section{SOMPZ+WZ: incorporating clustering information}
\label{sec:redshift_calibration_pipeline}
Using the SOMPZ method and including the previously described redshift calibration uncertainty, we obtain $10^8$ realizations of redshift distributions. We then proceed to use clustering information to further constrain the redshift realizations, with the overall methodology described in Section~\ref{sec:WZ} and the process of importance sampling on SOMPZ redshift realizations detailed in Section~\ref{sec:importance}.

\begin{figure}
    \includegraphics[width=\linewidth]{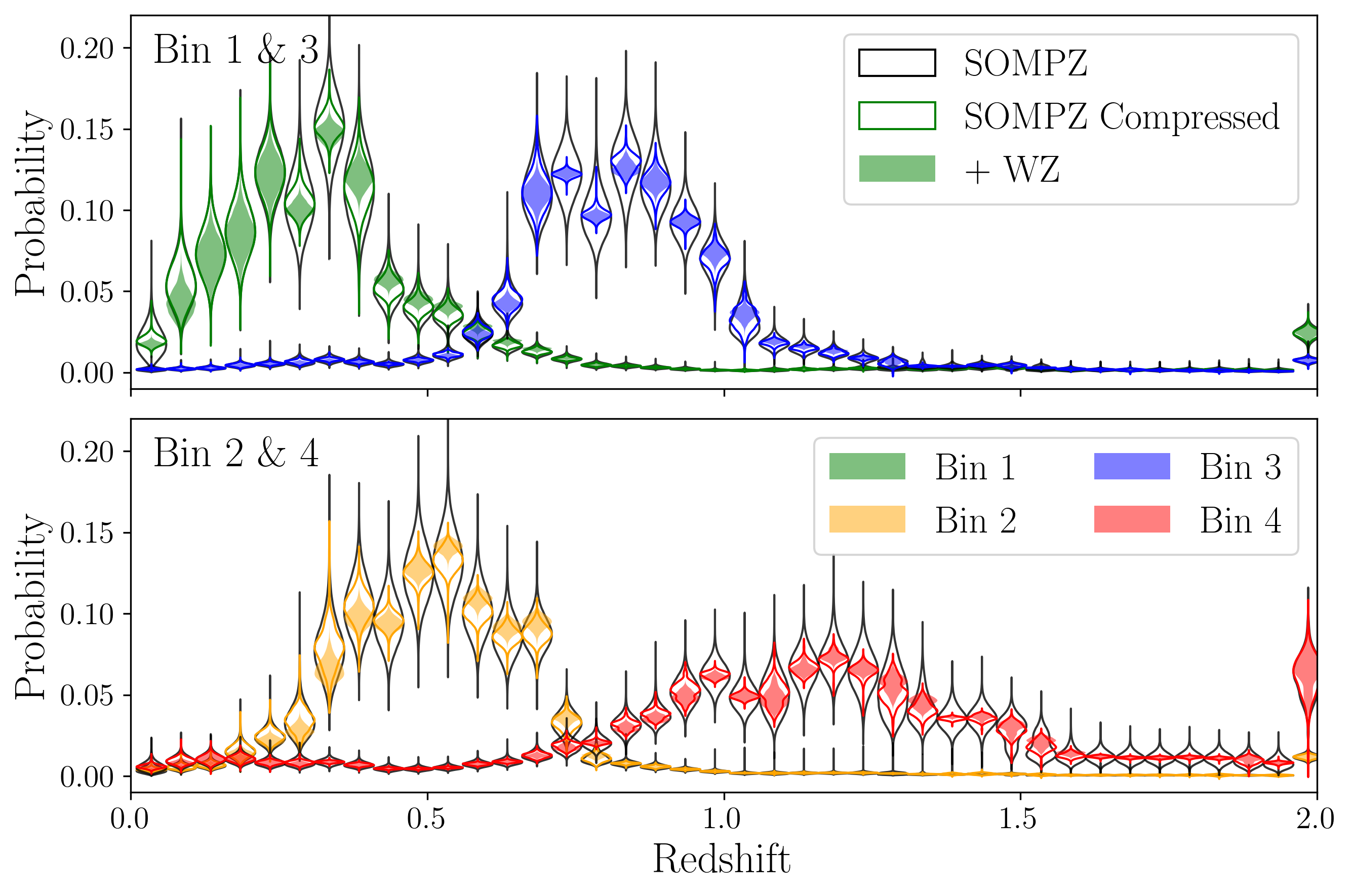}
    \caption{Violin plots showing the distribution of $n(z)$ functions in each of the four \mdet\ redshift bins. Black violin lines are the distributions of $n(z)$ taken directly from the SOMPZ redshift realizations. Coloured violin lines are the distributions after noise filtering with mode compression. Coloured solid violins further importance sample with the WZ likelihood. For the purpose of this figure, the points at $z=2$ hold the integral of $n(z)$ for $z\ge2$.} 
    \label{fig:violins2}
\end{figure}
\subsection{WZ methodology}
\label{sec:WZ}

In addition to SOMPZ, we can use information from galaxy clustering on small scales that is not used in the primary $3\times2$ observables \citep{y6-modeling}, following the DES-Y3 methodology presented in \citet{Gatti_Giulia_DESY3} and \citet{DESY3_MAGLIM_z}. We make use of the reference samples, denoted as the subscript $\rm r$, from spectroscopic BOSS-eBOSS surveys \citep{BOSS_color,eboss_dawson} distributed into small bins indexed by $z_i$, and cross-correlate them with the unknown sample $\rm u$, here the wide-field source sample distributed into to tomographic bins $b$, to get $w_{\rm ur}$. Each source galaxy is weighted by its response+statistical weight -- the same weighting scheme used in SOMPZ -- in both the model and measurements.
The model is written as $\hat{w}_{\rm ur}\left(b_{\rm u},\,b_{\rm r},\, \alpha_{\rm u},\,\alpha_{\rm r}, \,n_{\rm r}(z) ,\, n_{\rm u}(z) ,\,\{s^{\rm k}_{u}\} \right)$, which depends on the galaxy biases $b_{x}$, the magnification coefficients $\alpha_{x}$, the redshift distributions $n_{x}(z)$, and a set of nuisance parameters $\{s^{\rm k}_{u}\}$ for $k$-th Legendre polynomial. These nuisance parameters characterize potential systematics not incorporated in the model, such as the redshift evolution of $b_{\rm u}$ \citep{Gatti_Giulia_DESY3,Euclid_dassignies,y6-wz}.  
One can then introduce the likelihood of the measurements,
\begin{equation}
\label{eqn:wz_measurement}
\resizebox{.91\hsize}{!}{$
\mathcal{L}\left(w_{\text{ur}} | b_{\rm u}, b_{\rm r}, n_{\rm r}, n_{\rm u}, \{s_k^{\text{u}}\}\right) 
\propto \exp\left(-(w_{\text{ur}} - \hat{w}_{\text{ur}})^\top C_w^{-1} (w_{\text{ur}} - \hat{w}_{\text{ur}})/2\right)
$},
\end{equation}
where $C_w$ is the covariance of the cross-correlation. Using Bayes' theorem, and marginalizing over the nuisance parameters, one can then evaluate the cross-correlation likelihood of a redshift distribution $n_{\rm u}(z)$:
\begin{equation}
\label{eqn:wz_nz}
\mathcal{L} \left( w_{\rm ur} \! \mid \! n_{\rm u},  n_{\rm r} \right) \propto \int {\rm d} {\mathbf q}\, p({\mathbf q})\, \mathcal{L} \left( w_{\rm ur} \! \mid \! {n_{\rm u}, n_{\rm r}, \mathbf q} \right)\,
\end{equation}
with the parameter vector $\mathbf q = \{b_{\rm u},\,b_{\rm r}, \,\alpha_{\rm u},\,\alpha_{\rm r},\,\{s^{\rm k}_{u}\}\}$.

\begin{figure*}
    \centering
    \includegraphics[width=0.8\textwidth]{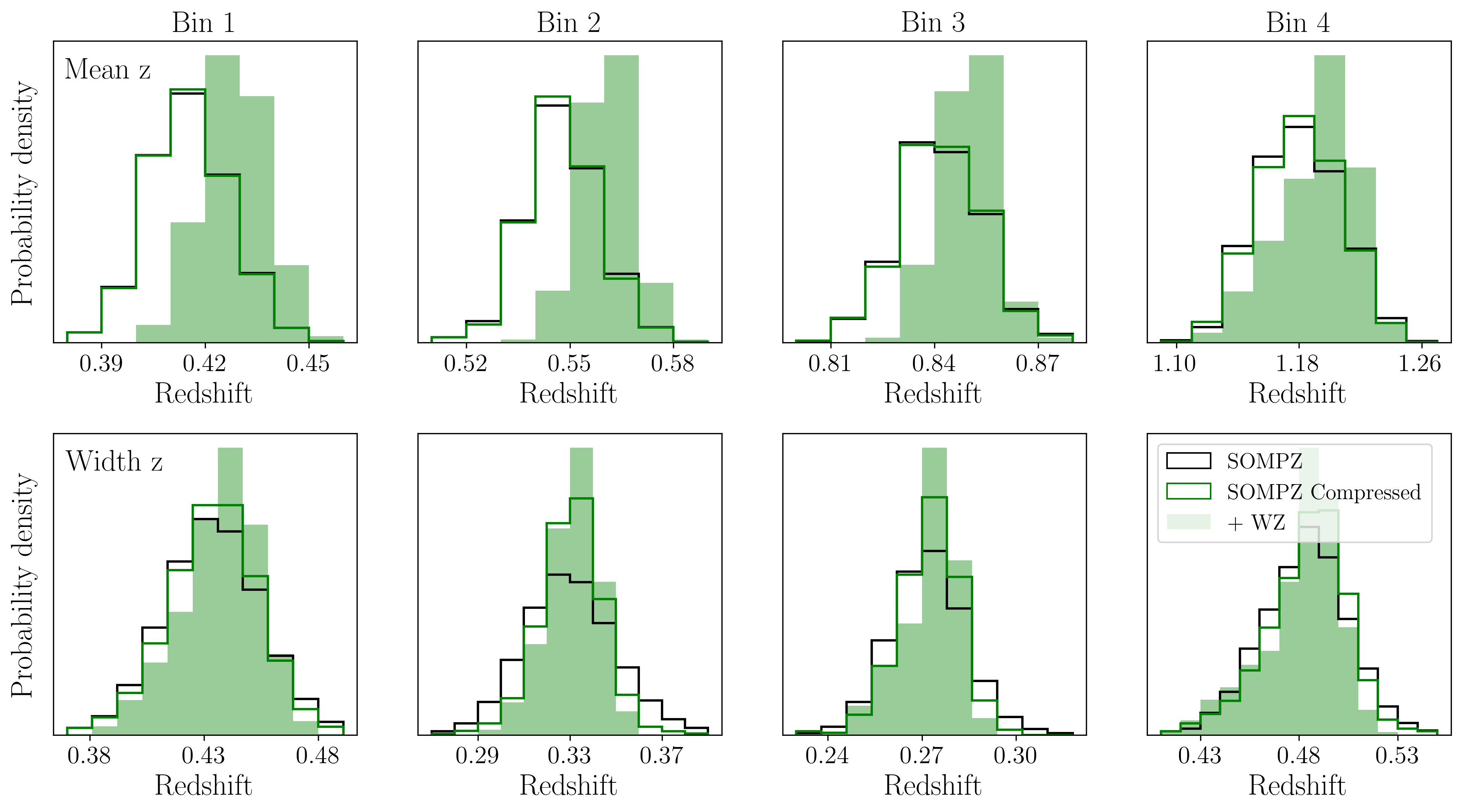}
    \caption{For each of the source redshift bins, we plot the distribution of the mean of $n(z)$ in the top row, and the distribution of the standard deviation of redshifts ${\rm std}(z)$ in the bottom row. The solid / dashed histogram indicates whether the $n(z)$ probability is determined using only photometric redshift data (SOMPZ), or the combination of SOMPZ with cross-correlation against eBOSS galaxies and quasars (SOMPZ$+$WZ).  The black / green colour is before / after the mode-projection technique has been applied to remove cosmologically irrelevant fluctuations in $n(z)$.  The clearest features are the upward $\approx 1\sigma$ shifts in $\bar z$ induced by adding WZ information, and some narrowing of the standard deviation distributions by the compression process.}
    \label{fig:meanz}
\end{figure*}

\begin{figure*}
    \includegraphics[width=0.8\textwidth]{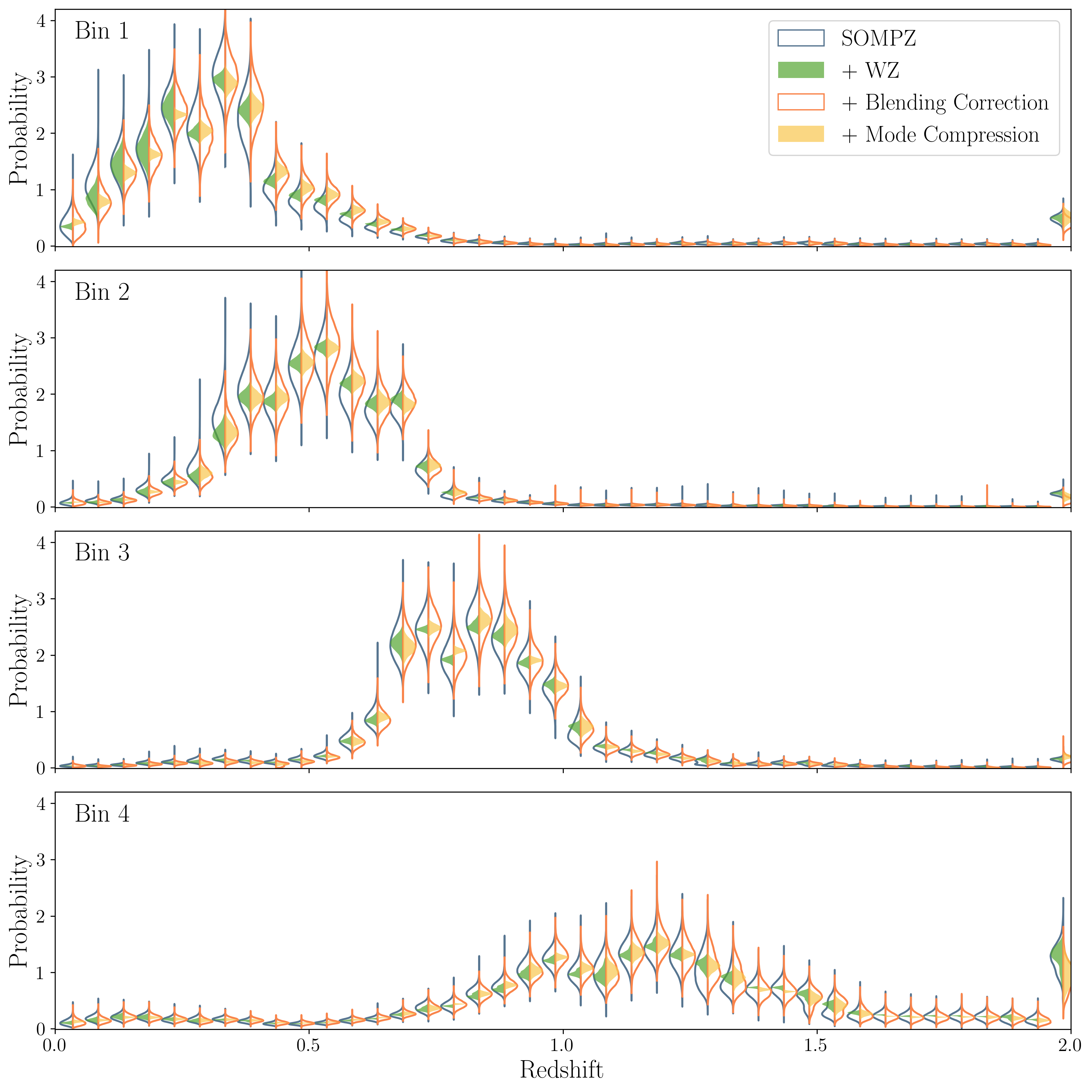}
    \caption{Violin plots showing the distribution of $n(z)$ functions in each of the four \mdet\ redshift bins.  Blue open violins are the distributions of $n(z)$ taken directly from the SOMPZ redshift realizations. Green filled violins are the distributions after importance sample with the WZ likelihood. Pink open violins further correct for the shear and redshift-dependent blending effect using image simulations. Orange filled violins have used mode projection to remove fluctuations with no significant cosmological impact (the mean is unchanged by this process) from blending corrected realizations. For the purpose of this depiction, the points at $z=2$ hold the integral of $n(z)$ for $z\ge2$. In our analysis, n(z) is piled up $z\ge3$ in the last step when we do mode compression. (Appendix~\ref{app:pileup}) }
    \label{fig:violins plot}
\end{figure*}

\begin{figure*}
    \centering
    \includegraphics[width=\textwidth]{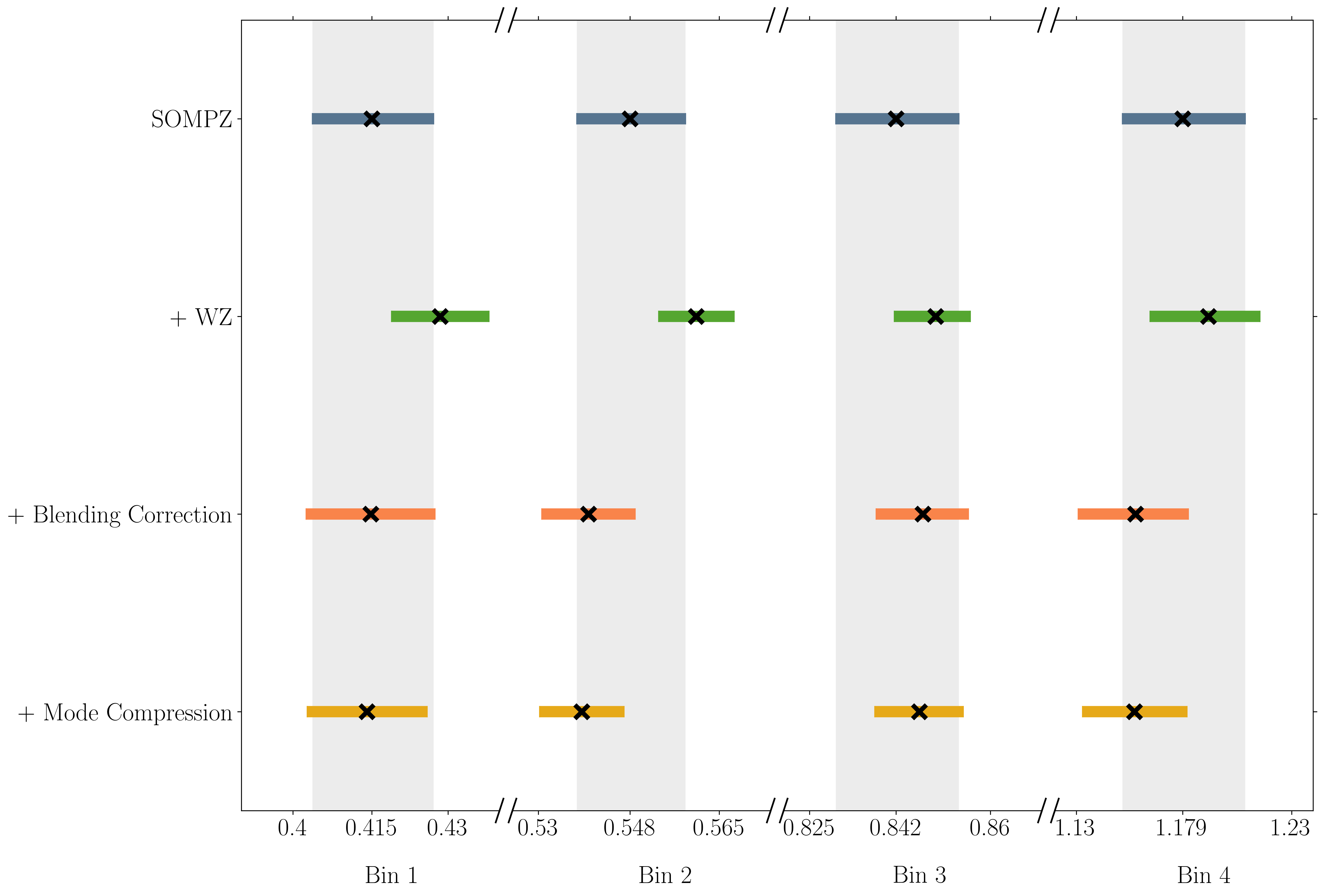}
    \caption{\label{fig:uncertainty_all} The redshift calibration pipeline proceeds in four steps: (1) SOMPZ redshift calibration with associated uncertainty, (2) adding clustering information to constrain the redshift distribution, (3) correcting for blending, and (4) applying mode compression to remove non-informative redshift modes. The mode-compressed outputs are used for cosmological inference. WZ shifts the mean redshifts higher, while blending shifts them lower. Mode compression is shown to align well with the pre-mode sampled blending-corrected distributions.
} 
\end{figure*}

\subsection{Noise filtering and WZ importance sampling}
\label{sec:importance}

We calculate the WZ likelihood $\likeli(\rm WZ | \vecn)$ of the cross-correlation data between eBOSS spectroscopic galaxies and the weighted \mdet\ source galaxies for each of the $10^8$ $n(z)$ realizations produced from incorporating redshift uncertainties.  Each $n(z)$ realization is a vector $\vecn$ of length 320 (4 bins, each with $n(z)$ tabulated at redshift intervals $\Delta_z=0.05$  from $0<z<4$).  \citet{y6-modes} describe an algorithm to derive encoder and decoder matrices $E$ and $D$ that can compress $\vecn$ into an $M$-dimensional $\vecu = E (\vecn- \bar\vecn),$ and reconstruct using $\hat\vecn=D\vecu + \bar\vecn.$  This is effectively a denoising and lossy compression operation on the redshift realizations: the process removes linear combinations of $\vecn$ that do not correspond to significant variations in the predicted observables $\vecc = \{\xi_\pm, \gamma_t\}.$ 
"Significance" is quantified here by first linearizing the model $\hat\vecc(\Omega_0,\vecn)$ as a function of deviations of $\vecn$ from the sample mean $\bar\vecn$, holding the cosmological parameters and other nuisance parameters fixed at nominal values $\Omega_0$.  For a given $n(z)$ sample $\vecn,$ we can define
\begin{equation}
    \chi^2 = \left[\vecc(\vecn) - \vecc(\hat\vecn)\right]^T C_c^{-1} \left[\vecc(\vecn) - \vecc(\hat\vecn)\right]
    \label{eq:chi2}
\end{equation}
with $C_c$ being the covariance matrix of the summary statistics in the DES Y6 analysis, derived in \citet{y6-modeling}.  The matrices $\matD$ and $\matE$ are optimized to have the smallest rank $M$ such that the average $\langle\chi^2\rangle$ over the $n(z)$ samples is below some chosen threshold.  At this stage, we choose $\langle\chi^2\rangle<0.05,$ for which all except $M=11$ linear modes of variation of $\vecn$ can be discarded as irrelevant noise.
Fig.~\ref{fig:violins2} shows the reduction in scatter in the violins between the original SOMPZ redshift realizations and the $\hat\vecn$ reconstructed after compression.  The mean $n(z)$ is, by construction, unchanged.

\begin{figure*}
    \centering
    \includegraphics[width=0.9\textwidth]{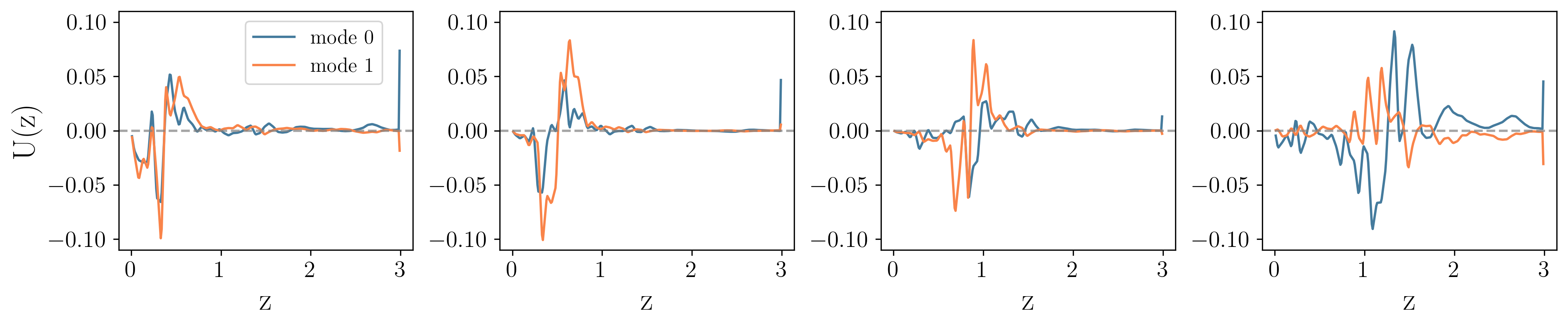}
    \caption{\label{fig:modes}
    The basis functions $U_{ij}(z)$ of $8\times10^4$ redshift realizations using SOMPZ + WZ + blending, shown for each of the four tomographic bins. The first and second of the seven modes are plotted. These $U_{ij}(z)$ represent the dimensional reduction of $\delta n(z)$s that impact cosmological inference. The jump in the tail at z = 3 arise from the pile up of redshift realizations at z=3. The full 7 modes are shown in Appendix \ref{app:modes}.}
\end{figure*}

This noise reduction in the redshift realizations conserves the probability distribution of cosmological impacts from $n(z)$ variations, while making the WZ likelihood distributions of the redshift realizations tighter with fewer spurious fluctuations. We next proceed by calculating $\likeli(\rm WZ | \hat\vecn)$ for each redshift realization, and retaining each realization with a probability proportional to $\likeli$ to yield a sample from the joint posterior of both the SOMPZ and WZ data, appropriately marginalized over the systematic uncertainties in both methods.  The constant of proportionality is chosen so as to yield $10^4$ samples from the SOMPZ$+$WZ posterior. As plotted in Fig.~\ref{fig:violins2}, the scatter in the violins are further reduced.

Fig.~\ref{fig:meanz} plots the distributions of the mean $z$ and standard deviation in $z$ of each bin's $n(z)$ function, where smaller standard deviations indicate more sharply peaked $n(z)$. These distributions are shown for the three cases:  original redshift realization samples, compressed redshift realization samples, and SOMPZ$+$WZ samples. The mean redshift $\bar z$ in each bin is shifted upwards by roughly the $1\sigma$ uncertainty in $\bar z$ by the addition of WZ information. \citet{y6-wz} show that the correlation functions of \mdet\ sources with eBOSS galaxies clearly favor higher $\bar z$ values than the mean $n(z)$ of the SOMPZ samples. Closer investigation reveals that this shift comes about in large part because the WZ data favor redshift realizations from those points in the Latin hypercube that have perturbed the \texttt{COSMOS2020} and \texttt{PAUS} redshifts upwards, with consistent results found when adopting broader WZ systematic priors. 

In the final step we define each SOMPZ$+$WZ $n(z)$ as a continuous function of $z.$ This is done by linear interpolating between the $n(z)$ at discrete values of separation $\Delta_z=0.05$, and rebinning to $\Delta_z = 0.01$. In other words, each discrete probability in $n(z)$ at value $z_i$ is spread into a function that rises linearly from zero at $z_i- \Delta_z$ to a peak at $z_i,$ then falls linearly to zero at $z_i+\Delta_z$, like a triangle. The only exception is the first element of $n(z),$ with $z_1=[-0.005,0.005]$: This probability is distributed into an asymmetric triangle that begins at $z=0.$  This enforces the condition that $n(z)\rightarrow 0$ at $z\rightarrow0,$ which is both physically required and mathematically required to avoid divergence in the calculation of observables. See Appendix \ref{app:triangular_binning} for more details.

\section{Correcting the n(z) for Blending effects}
\label{sec:blending}
Adding clustering redshift information to SOMPZ selects $10^4$ realizations of redshift distributions. We then proceed to correct for the line-of-sight galaxy blending with the overall methodology described in Section~\ref{sec:bl_methodology}, and perform the following steps on $n(z)$ realizations before sampling in the cosmological likelihood analyses:

\begin{enumerate}[label=\arabic*), align=left, leftmargin=*, labelwidth=2em]
 \item Correct for redshift- and shear-dependent blending effects in redshift calibration (Section~\ref{sec:final_nz})
 \item  Apply mode compression to reduce the dimensionality of the redshift realizations (Section~\ref{sec:modes})
\end{enumerate}

\subsection{Blending correction methodology}
\label{sec:bl_methodology}
When galaxies are observed in a photometric survey, their true three-dimensional position distribution is projected onto the two-dimensional sky. Through this effect, the light of galaxies that are distantly separated along the line of sight can become blended together on the observational plane. This results in the convolution of the redshift distribution and the shear response of the galaxy sample; specifically, blending results in a redshift-dependent shear bias, thereby sourcing and correlating the associated systematic uncertainties \citep{y3-blending}.
The correct redshift distribution that accounts for the response of weak lensing tracers to shear, and therefore, is appropriate to use for measurements of cosmic shear, is the effective redshift distribution for lensing, defined through the response+statistical weight. 

While the blending effect can not be directly measured from data, it is possible to quantify this effect using a set of image simulations for which the true response can be measured. For each set of simulations, we calibrate the response weighted redshift distribution $n^{\mathrm{imsim}}_i(z)$ using the same SOMPZ methodology as presented in the preceding sections for the four tomographic bins. Our goal is to construct a bias model $F_i(z)$ that corrects for blending effects, defined as
\begin{equation}
    \tilde{n}^{\mathrm{imsim}}_{i}(z) = \left(1 + F_i(z)\right) n^{\mathrm{imsim}}_{\mathrm{i}}(z),
\end{equation}
where $i$ represents the tomographic bin, and $n^{\mathrm{imsim}}_{\mathrm{i}}(z)$ is weighted by the response+statistical weight measured from \mdet on image simulation. In brief, the fitting procedure is to adjust $F_i(z)$ such that the integral over the effective redshift distribution $\tilde{n}^{\mathrm{imsim}}_{i}(z)$ matches the shear measured using \mdet over the true applied shear, for all 12 suites of image simulations. Further details for the $F_i(z)$ fitting procedure and also the SOMPZ pipeline for image simulations are described in \citet{y6-imagesims}.

This $F_i(z)$ is what we apply to our SOMPZ$+$WZ $n(z)$ realizations to correct for blending. For each of the $10^4$ SOMPZ+WZ $n(z)$ realizations, we draw 8 Markov chain Monte Carlo (MCMC) samples incorporating simulation imperfections from the bias model $F_i(z)$, and apply the correction as
\begin{equation}
    \tilde{n}^{\mathrm{SOMPZ+WZ+BL}}_{i}(z) = \left(1 + F_i(z)\right) n^{\mathrm{SOMPZ+WZ}}_{\mathrm{i}}(z).
\end{equation}
On the other hand, the multiplicative bias $m$ is defined as 
\begin{equation}
    m_i = \int \mathrm{d}z \, \tilde{n}^{\mathrm{SOMPZ+WZ+BL}}_{i} - 1
\end{equation}
for each redshift realization. Each redshift realization is further normalized so that its integral equals 1.

\subsection{Final $n(z)$}
\label{sec:final_nz}

The result of adding blending correction is plotted in Fig.~\ref{fig:violins plot}. The violins indicating the $n(z)$ distributions at the different $z$ show good agreement with each other. The SOMPZ$+$WZ (green filled) $n(z)$ violins are narrower than the SOMPZ (blue open) ones in each $z$ bin, demonstrating the value of clustering redshift in terms of its constraining power. The spread in redshift violins gets larger when propagating the blending correction into our SOMPZ$+$WZ $n(z)$ realizations, which includes uncertainty due to simulation imperfections.

The mean redshifts for these $n(z)$ realizations in each stage are shown in Fig.~\ref{fig:uncertainty_all}, with the error bars indicating $1\sigma$ uncertainty ranges and the cross symbols showing median redshift values. The corresponding numerical values are listed in Table~\ref{tab:uncertainty}. The WZ likelihood importance sampling selects SOMPZ redshift realizations, reducing the uncertainty from sample variance, shot noise, deep field calibration, and the redshift sample uncertainty. In particular, this selection leads to an increase in the mean redshift of each tomographic bin, as the  WZ likelihood preferentially selects redshift realizations with a higher \texttt{PAUS}/\texttt{COSMOS2020} redshift.

In DES Y6, due to using shear-dependent blending correction response measurement \mdet, we have a reduction of multiplicative shear bias $m$ compared to Y3. Meanwhile, we find a larger shift of the mean redshifts toward lower values from blending correction than in Y3. This mostly comes from that our Y6 sample is one magnitude fainter than Y3, and in part from chromatic PSF effects detailed in \citep{y6-imagesims}.

\subsection{Mode compression, sampling, and reconstruction} \label{sec:modes}

The $8 \times 10^4$ $n(z)$ realizations of the posterior of SOMPZ$+$WZ$+$BL are once again subjected to the mode-projection algorithm of \citet{y6-modes}, with the goal of dimensionality reduction to enable practical sampling as part of the cosmological Markov chains.  
This time we select $M=7$ modes, reducing the $\langle \chi^2\rangle$ attributable to compression losses to be below 0.15 (see Appendix~\ref{app:modes} for $\langle \chi^2\rangle$ vs $M$ analysis leading to choice of $M = 7$ modes). Fig.~\ref{fig:modes} shows the basis functions $U_{ij}(z)$ of the first and second modes, i.e.\ the portion of $j$th row of the decompression matrix $\matD$ that specifies the mode in redshift bin $i$. 
The adopted model for the $n_i(z)$ functions for bin $i$ is
\begin{equation}
 n_{i}(z) = \bar{n}_i(z) + \sum_{j=1}^M u_{j} U_{ij}(z) \; .
    \label{eq:uU}
\end{equation}
Because the SOMPZ$+$WZ$+$BL samples generate $\vecu$ values that are independent unit-normal distributions for each component, we adopt this simple distribution as our prior for cosmological analyses. After blending correction, shear and redshift are correlated. As a result, $\vecm$ and $\vecu$ are jointly sampled in MCMC chains based on their correlation matrix.

Fig.~\ref{fig:violins plot} further shows these reconstructed $n_i(z)$ functions used for DES Y6 cosmological analyses as the yellow shaded violins. The resulting space of $n(z)$ functions spans all cosmologically relevant fluctuations allowed by the source-galaxy photometry in wide and deep fields, redshift survey data, cross-correlation with eBOSS galaxies, and blending correction. In Fig.~\ref{fig:uncertainty_all}, one can see that the distributions are typically slightly narrower after the compression process than before. This should not be taken to mean that the compression has somehow shifted or altered the cosmological consequences of the source redshift distribution, because the mode compression is explicitly formulated to avoid changing the predicted distribution of cosmological statistics at observable levels.  In contrast, this difference indicates shortcomings of compressing $n(z)$ variations into means and standard deviations. For further validation on this conclusion, see Appendix~\ref{app:modes}.

\section{Validation and impact of redshift calibration choices on cosmology}
\label{sec:results_and_discussion}
In this section, we cross-check the results on our redshift calibration using shear ratios (Section \ref{sec:Shear_Ratio}), and validate our methodology using simulated 1$\times$2pt and 3$\times$2pt analyses (Section \ref{sec:consistency}).

\subsection{Shear Ratio}
\label{sec:Shear_Ratio}

A valuable cross-check in the redshift characterization process and a complementary source of information from galaxy-galaxy lensing is the shear-ratio (SR) test. By taking the ratio of two galaxy-galaxy lensing signals using the same lens bin but different source bins, this observable largely cancels out the dependence on the galaxy-matter power spectrum while preserving sensitivity to the angular diameter distances of the lens and source populations. In the absence of overlapping objects in source and lens bins, this observable is approximately scale independent, a limit that we refer to as "geometric". The SR dependence on angular diameter distances is weakly sensitive to cosmological parameters, leading to a dominant dependency on photometric redshift deviations.

To ensure the information is not affected by intrinsic alignment and other nuisance parameters, we focus on the two lens-source bin pairs which have the least lens-source overlap (see \citealt{y6-ggl} for further details), namely the two highest redshift source redshift bins and the two lowest redshift lens bins. The shear ratios we obtain are only sensitive to $n(z)$'s and shear calibration at measurable levels.

The SR is sensitive to the source redshift distributions, particularly to anti-correlated errors in the mean $z$ of source bins. The SR provides complementary information to SOMPZ and WZ, because it uses shear information rather than photometry, as in SOMPZ, or positions, as in WZ. Additionally, we use a maximum scale of $6 \, \text{Mpch}^{-1}$ to measure SR, which is the minimum scale applied to the galaxy-galaxy lensing analysis. The SR measurement hence has little correlation with the cosmic shear or the $3\times2$pt data vector. Note that the second lens bin is later excluded from the cosmological analysis due to the poor fit of theoretical prediction to the data. Since the SR test here is purely geometric and restricted to small scales, it is not affected if the problem arises from unknown systematics in the density or shear tracer, such as imaging systematics affecting the density weights for clustering. The SR result will only be affected if the poor fit in lens bin 2 arises from its redshift calibration. To ensure robustness against this possibility, we performed two additional tests: one applying a flat prior on the lens bin 2 mean source redshift shift and the other using the SR which only includes the lens bin 1. The results are consistent with those obtained when including both lens bins 1 and 2, though losing approximately $14\%$ and $37\%$ of the constraining power, respectively.

\begin{figure}
    \centering
    \includegraphics[width=\columnwidth]{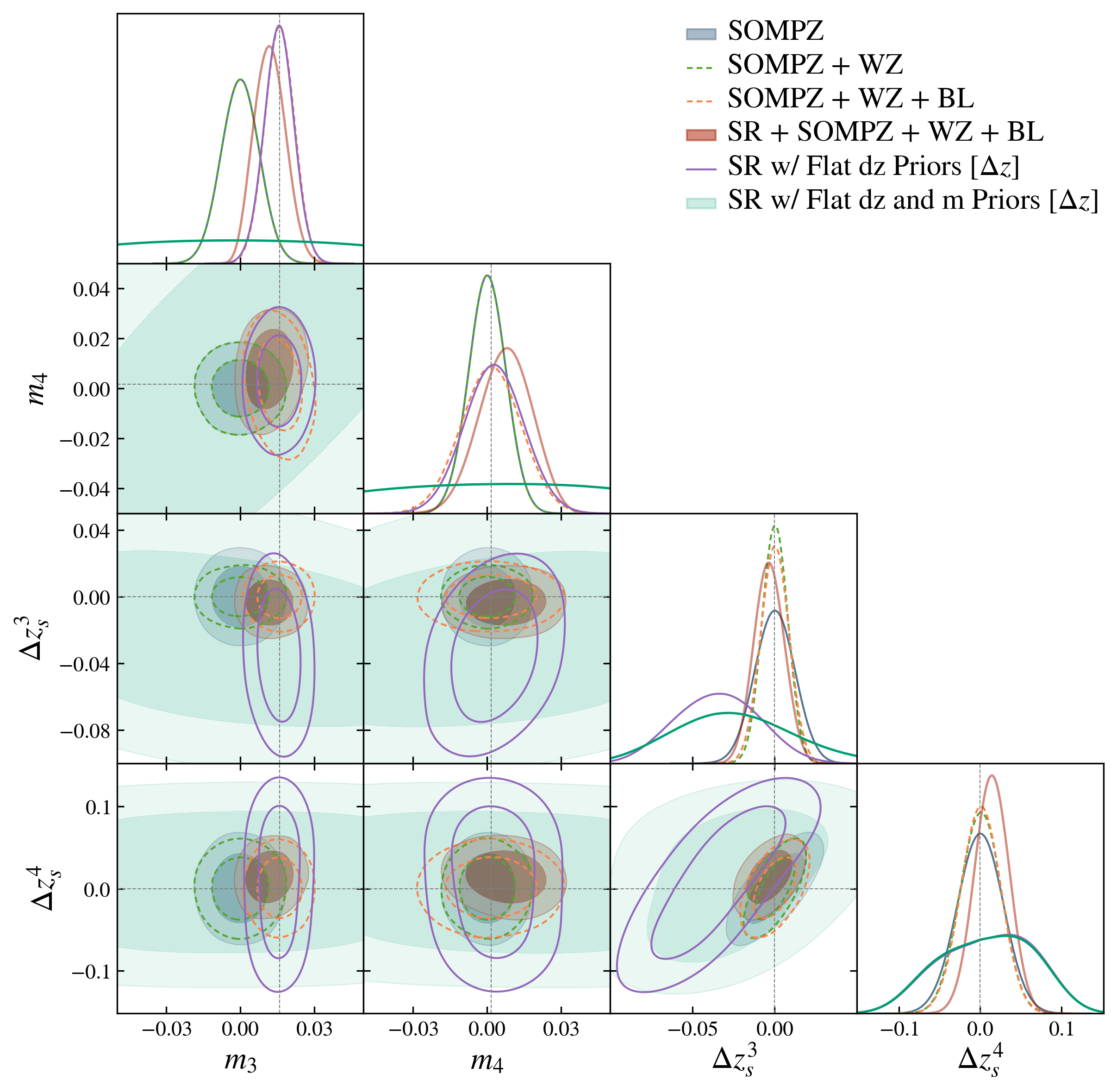}
    \caption{\label{fig:dz_SR} Mean source redshift and multiplicative shear bias constraints from the shear-ratio only chain for the two highest redshift source bins, with (i) flat priors $[-0.1,0.1]$ on $\Delta z_i$ only and (ii) flat priors on both $\Delta z_i$ and $m_i$, to compare it with the combination of the other redshift calibration method: (i) SOMPZ (ii) SOMPZ + WZ and (iii) SOMPZ + WZ + BL. The SR with flat priors on only $\Delta z_i$ absorbs the $\Delta z_i$–$m_i$ degeneracy, imposing a shift in $\Delta z_3$ and $\Delta z_4$. The SR with flat priors on both $\Delta z_i$ and $m_i$ is in agreement with the SOMPZ+WZ, validating the calibration methods used in this work.
}
    
\end{figure}

Using the unblinded DES Y6 small-scale galaxy-galaxy lensing data vector, we use MCMC chains to constrain 4 parameters quantifying systematic errors. We model errors in the $n_i(z)$ distribution of bin $i$ with shifts $\Delta z_i$ such that $n_i(z) = \bar{n_i}(z + \Delta {z_i})$, where $\bar{n_i}$ is the mean SOMPZ+WZ+BL estimate -- such shifts in $n_i(z)$ alter the SR at first order. We also allow multiplicative shear biases $m_i$ for each source bin.  Fig.~\ref{fig:dz_SR} shows the posterior distributions of these parameters given the likelihood of the measurements and two different priors: (i) flat priors on $\Delta z_i$ only and (ii) flat priors on both $\Delta z_i$ and $m_i$, with the straight dashed line marking the SOMPZ+WZ+BL central values. The posterior constraints from SR data are compared to the ranges of uncertainty of $\Delta z_i$ propagated from each of the three redshift estimations: (i) SOMPZ, (ii) SOMPZ+WZ, and (iii) SOMPZ+WZ+BL. For each of the three cases, we generate $10^4$ realizations of $n(z)$ for each source redshift bin from their corresponding modes, using equation~(\ref{eq:uU}). From these $n(z)$ realizations, we calculate the shifts $\Delta z_i$ in the mean of $n(z)$.  Following the same methodology, we can also constrain the $\Delta z_i$ and $m_i$ values with the combined information of SR+SOMPZ+WZ+BL data.

The agreement of the green SR contours in Fig.~\ref{fig:dz_SR} with the dashed red SOMPZ+WZ+BL contours verifies the consistency of the independent SR measurements with $n(z)$ estimates derived in this paper. Thus, the shear-ratio test is passed, confirming the validity of the model for the two highest redshift source bins.

\begin{figure*}
     \centering
    \includegraphics[width=0.9\textwidth]{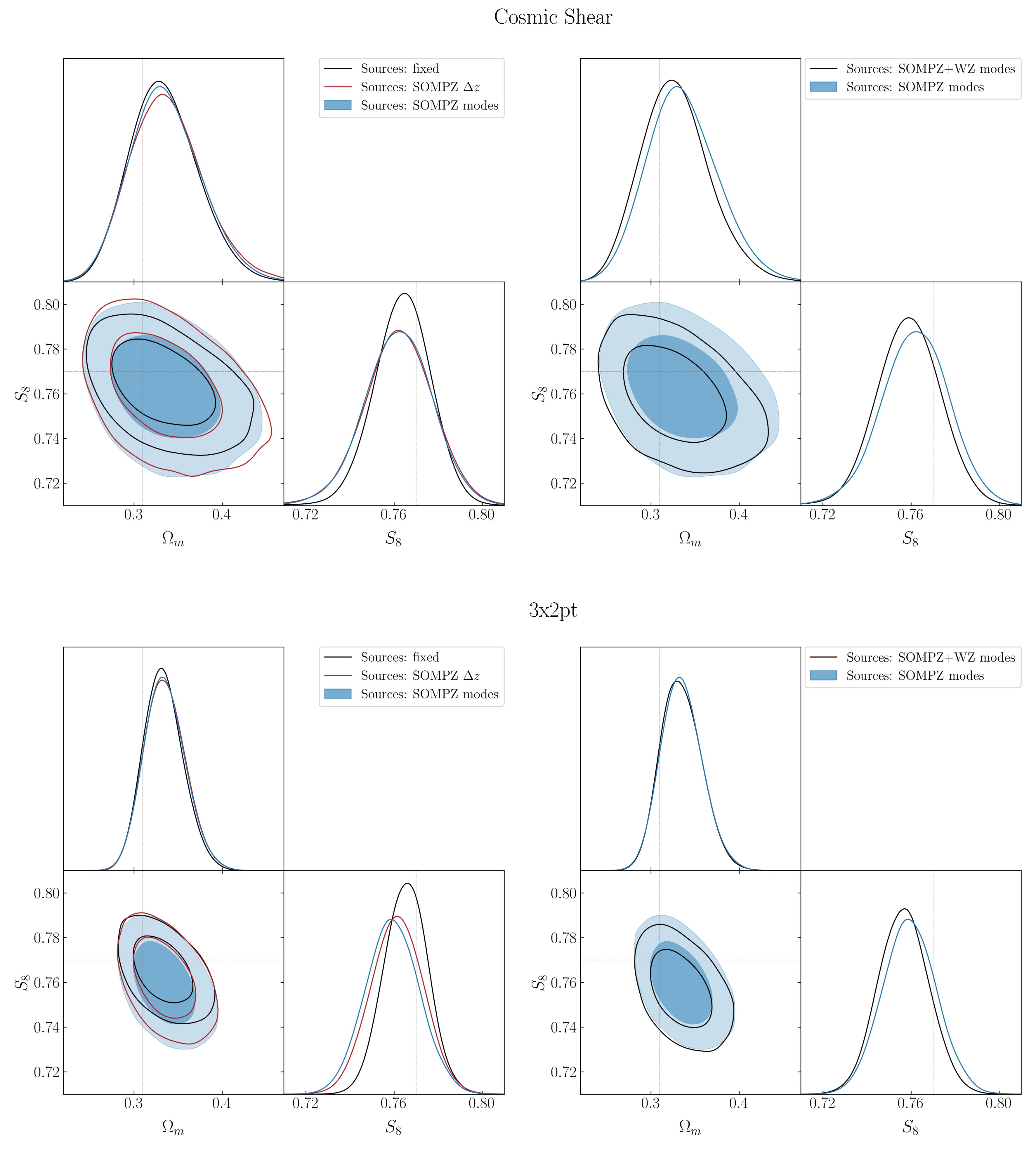}
     \caption{\label{fig:combined_chain_plot} Impact of photometric redshift analysis choices on cosmological parameter constraints on $\Omega_m$ and $S_8$ for the cosmic shear (upper panels) and $3\times2$pt (lower panels) analysis. \textit{Left:} The impact of the redshift uncertainty modeling using a fixed redshift distribution, mean redshift shift, or modes to capture the full uncertainty. \textit{Right:} The impact of combining clustering redshifts with the SOMPZ information. In the 3$\times$2pt chains, we use WZ modes for lenses, and marginalize over the redshift uncertainty of both lenses and sources. The less informative priors shift and broaden the posteriors, highlighting the impact of redshift uncertainties on parameter constraints.}
\end{figure*}

\subsection{Consistency of independent methods} 
\label{sec:consistency}
In this section, we validate the consistency of independent methods for photometry redshift calibration in cosmological inference. We use simulated datavectors generated by \cite{y6-modeling}, where the source and lens redshift used in the simulation is SOMPZ+WZ.

We compare the impact of photometric redshift analysis choices on cosmological parameter constraints ($\Omega_m$, and $S_8$) for the cosmic shear (upper panels) and $3\times2$pt (lower panels) analysis in Fig.~\ref{fig:combined_chain_plot} \citep{cosmosis, getdist}. On the left, we compare the impact of redshift uncertainty on the $\Omega_m - S_8$ constraint. We show three cases: using a fixed redshift distribution for the sources, i.e. neglecting redshift calibration uncertainty, using first order redshift distribution uncertainty $\Delta z$, and the mode sampling method that preserves all shape perturbation and correlation between redshift bins. The fixed $n(z)$ yields the tightest constraints. This is expected since we assume perfectly known redshift distribution with no uncertainty. SOMPZ $\Delta z$ and SOMPZ modes give very similar constraints (0.004$\sigma$ for 1$\times$2pt, and 0.039$\sigma$ for 3$\times$2pt), with modes yielding a slightly tighter constraint in 1x2pt and a slightly looser constraint in 3$\times$2pt. Compared to the mode compression which uses all cosmology relevant variation in $n(z),$ compressing the allowed variation of $n(z)$ into a single $\Delta z$ parameter per bin could overestimate cosmological accuracy by suppressing some variation in $n(z)$ that is covariant with cosmology. It could also, however, underestimate cosmological accuracy if the compression rectifies cosmologically irrelevant noise in $n(z)$ into shifts that do affect cosmology. We see that neither effect is large in this case, but the use of the modes is a more rigorous and  trustworthy compression.

On the right, we compare the impact of adding clustering redshift information to the SOMPZ estimated redshift distribution, in both cases using modes as the redshift distribution sampling method. The SOMPZ + WZ mildly shifts (0.085$\sigma$ for 1$\times$2pt, and 0.046$\sigma$ for 3$\times$2pt) and better constrains $\Omega_m-S_8$, tightening $S_8$ by $10\%$ for 1$\times$2pt and $7\%$ for 3$\times$2pt, showing the importance of WZ. 

Due to parameter degeneracies and projection effects, the $\Omega_m - S_8$ posterior is not centered on the true value used in generating the data vector (shown as the dashed line). This is more pronounced in the 3$\times$2pt case, where projection effects are amplified due to the marginalization over a higher-dimensional nuisance parameter space.

\section{Discussion \& Conclusion}
\label{sec:discussion}

\begin{figure}
    \centering
    \includegraphics[width=\columnwidth]{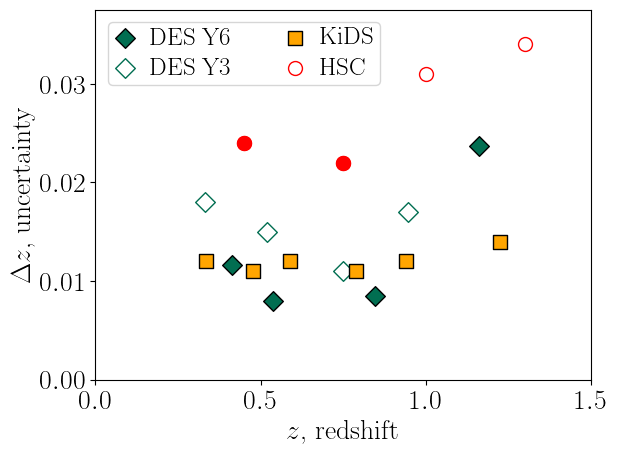}
    \caption{\label{fig:uncertainty_surveys}
    Uncertainty of redshift calibration for DES Y6 (green shaded), compared to DES Y3 \citep*{y3-wlpz} (green open), KiDs \citep{KiDSlegacy_redshift} (yellow), and HSC \citep{HSC_y3_redshift} (red). The two unfilled red circles are not used as informative priors in HSC. DES Y3 and Y6 are dominated by sample variance of the calibration fields at low redshift and redshift sample uncertainty from \texttt{COSMOS2020}/\texttt{PAUS} at high redshift. DES Y6 has smaller uncertainty than Y3 over similar redshift ranges, owing to improvements in the $g$ band, SOM algorithm, and calibration samples.}
    
\end{figure}

As the largest of the Stage III imaging surveys, DES has mapped 5000 deg$^2$ of sky area to provide an unprecedented dataset to probe the nature of dark matter and dark energy. During the $\sim$20 years since its inception, the experiment has driven the development of novel approaches for all stages of calibration, modeling, and analysis, including photometric redshift methodology, demanded by the rapidly improving statistical power and depth of the data. The final DES Y6 catalog ($i<24.7$) is more than a magnitude deeper than the Y3 data ($i<23.5$) and includes the $g$-band, owing to improvements in the colour-dependent PSF modeling and star-galaxy separation \cite{desy6_psf}. We summarize here the history of redshift calibration within the Dark Energy Survey, and the increasing complexity demanded by modern lensing analyses.

The \textbf{DES Science Verification} \citep{dessv_photoz} compared four redshift calibration methods: the fiducial choice, Skynet -- an artificial neural network with three hidden layers \citep{skynet_2014, skynet_2015}, ANNz2 -- randomized regression over 100 neural networks with varied configurations \citep{annz2}, TPZ -- random forest \citep{tpz}, and BPZ -- template fitting \citep{bpz}. Their performance was assessed in terms of the binned redshift distributions with spectroscopic validation data and the impact on the cosmological inference via simulated analyses. The uncertainty was determined as the spread in redshift of the four methods and propagated into the cosmology inference as a Gaussian prior on the mean redshift per bin. 

The \textbf{DES Year 1} analysis \citep{desy1_photoz} used the template fitting method BPZ \citep{bpz} for redshift calibration. The mean redshift bias was calibrated by matching to the \texttt{COSMOS2015} validation data, accounting for uncertainty due to limitations in validation data itself, which includes sample variance and photometric calibration uncertainty in \texttt{COSMOS2015}, as well as hidden variables and systematic errors from the matching process. Clustering redshift information was introduced as a cross-check, measured using \textsc{redmagic} galaxies, and in combination to yield a final constraint on the mean redshift. The uncertainty was propagated into the cosmology inference as a Gaussian prior on mean redshift per bin, with correlations of the mean-redshift uncertainty across tomographic bins accounted for by inflating the uncertainty with a conservatively estimated factor of 1.6. Further cross-checks were performed with the shear ratio method \citep{desy1_sr} and the DNF machine learning method \citep{dnf}. 

The \textbf{DES Year 3} analysis ($riz$+$ugJHK_{\rm s}$, $i<23.5$; \citet*{y3-wlpz}) introduced the SOMPZ method, an unsupervised machine learning approach that calibrates few-band wide optical data with multi-band deep optical+near-infrared photometry that has high-resolution, overlapping spectroscopic and narrow-band photometric redshift measurements. The uncertainties due to the limitation of spectroscopic redshifts and deep sample size, area, photometric calibration, and redshift accuracy were comprehensively modeled. With enhanced cosmological precision, the impact of the blending of galaxies on the shear bias and redshift distributions were jointly corrected using image simulations. Clustering redshifts \citep*{Gatti_Giulia_DESY3} and shear ratio tests \citep*{y3_sr} served as cross-validation and were both incorporated into the final redshift uncertainty.
The final ensemble of redshift realizations, where each represents a plausible instance given the uncertainty, contains information from the $n(z)$ shape uncertainty. Attempts to sample the redshift realizations were made using \textsc{Hyperrank} \citep{hyperrank}, which compressed the realizations onto a hypercube and smoothly sampled from it. However, the likelihood of the sampled redshift distribution is not continuous and induced a convergence problem during cosmological inference. As a result, the redshift realizations were primarily reduced to mean redshift perturbations in the cosmological inference. 

The \textbf{DES Year 6} analysis ($griz$+$uJHK_{\rm s}$, $i<24.7$) is built upon the SOMPZ framework with an improved unsupervised machine learning algorithm, \textsc{SOMF}, that has better performance on faint galaxies and significantly reduces bin overlap \citep{somf}. The uncertainty due to the imperfections in the redshift calibration sample, field-to-field variations in the photometric calibration of deep sample, and limitations in the size and area of both the deep and redshift calibration samples are modeled. We use a more realistic and conservative approach to account for the redshift calibration sample's imperfection that considers the bias in the narrow-band photometric redshift sample compared to the subset with observed spectra. We combine the colour-based SOMPZ $n(z)$ with clustering redshift information using measurements from eBOSS spectra \citep{y6-wz}. We correct for the impact of blending, now measured more precisely in redshift, using multi-band image simulations \citep{y6-imagesims}. Finally, we validate the resulting $n(z)$ using the shear ratio method \citep{y6-ggl}. To propagate the redshift realizations into the cosmological inference, we use a new mode sampling approach -- a modified PCA method that samples only the perturbations on the $n(z)$ shape that most influence the cosmology \citep{y6-modes}. 

Finally, we compare our calibration results with other stage III surveys, KiDS and HSC, in Fig~\ref{fig:uncertainty_surveys}, and to DES Y3 results. We adopt the fiducial setup for KiDs, using the SKiLLS-based SOM calibration with a conservative error floor of $\langle \delta z \rangle$ = 0.01 \citep{KiDSlegacy_redshift}. Our first three bins show significant improvements from Y3, due to the addition of $g$ band, the improved SOMF designed for faint galaxies, and the new \texttt{COSMOS2020} \citep{cosmos2020} calibration data to replace \texttt{COSMOS2015} \citep{cosmos2015}. The fourth bin shows a larger uncertainty, primarily due to the limited spectroscopic coverage at the fainter magnitude limit of Y6. The Stage III surveys show consistent uncertainty estimations on their redshift distributions.

We are now entering an era of Stage IV surveys. \citet{LSST_SRD} requires the mean redshift uncertainty of each source tomographic bin to not exceed $0.002(1 + z)$ in the Y1 DESC WL analysis and $0.001(1 + z)$ in Y10 DESC WL analysis, while \citet{Roman_SRD} similarly has a requirement of $0.002(1 + z)$ for the full High Latitude Imaging Survey, and \citet{euclid_srd} having the same requirement as in Roman. Investigations into whether these requirements can be met are underway (LSST \citealt{LSST_Graham,LSST_TQ, myles_prep}; Euclid \citealt{Roster_euclid,kang_prep}). To meet stringent requirements for future surveys, a deeper spectroscopic sample is essential to mitigate our current dominant contributor to redshift error. The Complete Calibration of the colour-Redshift Relation (C3R2) \cite{c3r2_2017,c3r2_2019,c3r2_2021}, with the goal of mapping the empirical relation between galaxy redshift, has already filled $84\%$ of their SOM in the range $0.2 < z < 2.6$. Their criteria for a SOM cell to be considered filled is the presence of at least one galaxy \citep{c3r2_2021}, which may not be statistically robust or complete for our use. DESI C3R2 (DC3R2) \citep{dc3r2} builds upon C3R2, significantly improving the C3R2 SOM statistics with more galaxies at lower redshift $0 < z < 1.6$, for $56\%$ of the C3R2 SOM. Possible future surveys like the Subaru-PFS/Roman (SuPR) Deep Survey \citep{supR2025}, 4MOST C3R2 (4C3R2) \citep{4c3r2}, and DESI/PFS support programs of Rubin will hopefully provide complete spectroscopic sample down to fainter magnitudes, with 4C3R2 spanning a large area to reduce sample variance, DESI showcasing its deep exposure capabilities \citep{dey_prep}, while SuPR, with its optical-through-infrared sensitivity, into $z > 1.5$ across a broad area. On the other hand, DESI \citep[e.g.,][]{DESI} has and will continue to provide millions of spectroscopic galaxies. However, samples at its higher redshift range, such as "luminous red galaxies" ($0.4 < z < 1.0$) and "emission line galaxies" ($0.6 < z < 1.6$), have stringent colour selections and cannot be used alone in a SOMPZ analysis such as this one.  They can, however, serve as much more powerful "reference" galaxy samples for WZ measurements, to the extent that they overlap a given source sample on the sky.

The development of new photometric redshift methods in the DES survey over the last decade has been instrumental to achieving its cosmological goals with weak lensing and large-scale structure. While significant spectroscopic resources will be required to go beyond the current state-of-the-art used in the DES analyses to reach the requirements for Stage IV photometric surveys, the work in DES and other Stage III surveys lays solid ground on which to build toward achieving those goals over the next decade.

\section*{Acknowledgements}
\textbf{Author Contributions:} All authors contributed to this paper and/or carried out infrastructure work that made this analysis possible. BY performed most of the analysis and manuscript preparation. AAm contributed to the analysis and manuscript preparation, prepared the redshift calibration sample, and developed the Y6 redshift calibration sample uncertainty method with BY. AC contributed to the implementation of the SOMF method. Troxel contributed with manuscript preparation. WA ran the clustering redshfit measurement and contributed to the manuscript preparation. GB and Troxel developed the modes pipeline. GB also performed the importance sampling and contributed to the manuscript preparation. GC ran the shear ratio consistency tests and contributed to the manuscript preparation. SM and MRB ran the blending correction and contributed to the manuscript preparation. GG and AAl contributed to analysis interpretation as conveners of the redshift Working Group (WG). GG and CC also contributed to the manuscript preparation on the script for Figure 1. DG and JMc contributed to manuscript preparation as collaboration internal reviewers. MY made the \mdet catalog and contributed to the manuscript preparation. DA made the \Balrog catalog and contributed to the manuscript preparation. SD contributed to the implementation of SOMF with AC. CS contributed to analysis interpretation as convener of the DES Y6 redshift WG at the time of analysis. JMy contributed to the analysis interpretation as DES Y3 source redshift calibration author. JP contributed to the manuscript preparation and to analysis interpretation as convener of weak lensing WG with AA. CC contributed to the analysis interpretation as Science Committee coordinator with Troxel. MC contributed to the analysis interpretation as Y6 analysis coordinator with MRB. The remaining authors have made contributions to this paper that include, but are not limited to, the construction of DECam and other aspects of data collection; data processing and calibration; calibration and source catalog creation; developing broadly used methods, codes, and simulations; running pipelines and validation tests; and promoting the science analysis.

Funding for the DES Projects has been provided by the U.S. Department of Energy, the U.S. National Science Foundation, the Ministry of Science and Education of Spain, 
the Science and Technology Facilities Council of the United Kingdom, the Higher Education Funding Council for England, the National Center for Supercomputing 
Applications at the University of Illinois at Urbana-Champaign, the Kavli Institute of Cosmological Physics at the University of Chicago, 
the Center for Cosmology and Astro-Particle Physics at the Ohio State University,
the Mitchell Institute for Fundamental Physics and Astronomy at Texas A\&M University, Financiadora de Estudos e Projetos, 
Funda{\c c}{\~a}o Carlos Chagas Filho de Amparo {\`a} Pesquisa do Estado do Rio de Janeiro, Conselho Nacional de Desenvolvimento Cient{\'i}fico e Tecnol{\'o}gico and 
the Minist{\'e}rio da Ci{\^e}ncia, Tecnologia e Inova{\c c}{\~a}o, the Deutsche Forschungsgemeinschaft and the Collaborating Institutions in the Dark Energy Survey.

The Collaborating Institutions are Argonne National Laboratory, the University of California at Santa Cruz, the University of Cambridge, Centro de Investigaciones Energ{\'e}ticas, 
Medioambientales y Tecnol{\'o}gicas-Madrid, the University of Chicago, University College London, the DES-Brazil Consortium, the University of Edinburgh, 
the Eidgen{\"o}ssische Technische Hochschule (ETH) Z{\"u}rich, 
Fermi National Accelerator Laboratory, the University of Illinois at Urbana-Champaign, the Institut de Ci{\`e}ncies de l'Espai (IEEC/CSIC), 
the Institut de F{\'i}sica d'Altes Energies, Lawrence Berkeley National Laboratory, the Ludwig-Maximilians Universit{\"a}t M{\"u}nchen and the associated Excellence Cluster Universe, 
the University of Michigan, the National Optical Astronomy Observatory, the University of Nottingham, The Ohio State University, the University of Pennsylvania, the University of Portsmouth, 
SLAC National Accelerator Laboratory, Stanford University, the University of Sussex, Texas A\&M University, and the OzDES Membership Consortium.

Based in part on observations at Cerro Tololo Inter-American Observatory at NSF's NOIRLab (NOIRLab Prop. ID 2012B-0001; PI: J. Frieman), which is managed by the Association of Universities for Research in Astronomy (AURA) under a cooperative agreement with the National Science Foundation.

The DES data management system is supported by the National Science Foundation under Grant Numbers AST-1138766 and AST-1536171. The DES participants from Spanish institutions are partially supported by MICINN under grants PID2021-123012, PID2021-128989 PID2022-141079, PID2023-153229NA-I00, SEV-2016-0588, CEX2020-001058-M and CEX2020-001007-S, some of which include ERDF funds from the European Union. IFAE is partially funded by the CERCA program of the Generalitat de Catalunya. The project that gave rise to these results received the support of a fellowship from “la Caixa” Foundation (ID 100010434). The fellowship code is LCF/BQ/PI23/11970028.

We acknowledge support from the Brazilian Instituto Nacional de Ci\^enciae Tecnologia (INCT) e-Universe (CNPq grant 465376/2014-2).

Research leading to these results has received funding from the European Research
Council under the European Union's Seventh Framework Program (FP7/2007-2013) including ERC grant agreements 240672, 291329, and 306478.

This document was prepared by the DES Collaboration using the resources of the Fermi National Accelerator Laboratory (Fermilab), a U.S. Department of Energy, Office of Science, Office of High Energy Physics HEP User Facility. Fermilab is managed by Fermi Forward Discovery Group, LLC, acting under Contract No. 89243024CSC000002.

\section{Data Availability}
The DES Y6 data products used in this work, as well as the full ensemble of DES Y6 source galaxy redshift distributions described by this work, are publicly available at https://des.ncsa.illinois.edu/releases. The cosmology likelihood sampling software we use \texttt{cosmosis} is available at https://github.com/joezuntz/cosmosis.


\input{mnras_template.bbl}



\newpage
\appendix

\section{SOMF details}\label{app:SOMF}
When placing a galaxy in the SOM, we select the cell whose flux is 'most similar' to the galaxy's flux. The distance metric is used to quantify this similarity, with smaller values indicating a better match. In the general SOM framework, this metric is a squared Euclidean distance $d(\mathbf{F}_i, \mathbf{F}_c) = \sum_{b=1}^{\text{flux}} \left( F_{ib} - F_{cb} \right)^2$
where $x_{ib}$ and $w_{cb}$ denote the $b$-band flux for galaxy $i$ and cell $c$ respectively. In our case, the metric is specifically designed for photometric redshift calibration, accounting for the properties of galaxy flux measurements and their uncertainty:
\begin{equation}
\label{eqn:distance_metric}
d(\mathbf{F}_i, \boldsymbol{\sigma}_i,\mathbf{F}_c) = \inf_{s} \left[ \tilde{d}(\mathbf{F}_i, \boldsymbol{\sigma}_i, e^s \mathbf{F}_c) + \frac{s^2}{\sigma_s^2} \right],
\end{equation}
where $\mathbf{F}_i$ is the multi-band flux for galaxy $i$, $\boldsymbol{\sigma}_i$ is the respective flux uncertainties, and $\mathbf{F}_c$ is the flux for cell c. To account for photometric noise, $e^s$ add fuzziness to cells, improving robustness to $S/N$ variations, while $\frac{s^2}{\sigma_s^2}$ penalizes excessive fuzziness. The $\text{inf}$ operator selects the optimal fuzziness factor s that minimizes total cost. The term $\tilde{d}$ is defined as:
\begin{equation}
\label{eqn:distance_metric_expansion}
\tilde{d}(\mathbf{F}_i, \boldsymbol{\sigma}_i,\mathbf{F}_c) = \sum_{b} \left[ \frac{\text{asinh}\, \nu_{cb} + W_{ib} \log 2\nu_{cb}}{1 + W_{ib}} - \text{asinh}\, \nu_{ib} \right]^2 \times \left(1 + \nu_{ib}^2\right),
\end{equation}
where we define the floored uncertainty of galaxy $s_{ib}$, the flux in units of signal-to-noise ratio $v_{ib}$ for galaxy $i$, and $v_{cb}$ for cell $c$ as:
\begin{equation}
\label{eqn:snr_flux}
s_{ib} \equiv \max \left( \sigma_{ib}, \frac{F_{ib}}{\text{SNR}_{\text{max}}} \right), \
\nu_{ib} \equiv \frac{F_{ib}}{s_{ib}}, \
\nu_{cb} \equiv \frac{F_{cb}}{s_{ib}},
\end{equation}
with the weight $W_{ib}= e^{2(\nu_{ib} - 4)}$ providing a smooth transition between the low and high $S/N$ regimes.

We detail our improvements over DES Y3 SOM here, and refer readers to \citet{somf} for further information:

1) SOMF uses the signal-to-noise, $S/N$, information for each galaxy, as defined in equation~(\ref{eqn:snr_flux}). This ensures that galaxies observed with higher $S/N$ have more influence on the metric. In addition, $S/N$ is used during the training phase. In the standard SOM implementation, cell flux is updated by a linear shift based on the difference between the input galaxy flux and the cell flux. In SOMF, this shift is determined by the $S/N$ level of the input galaxy: for flux bands with high $S/N$, the cell flux is shifted linearly, while for low $S/N$ bands, no update is applied. These $S/N$ treatments that penalizes low $S/N$ photometry during training while not overly reliant on high $S/N$ bands in the metric are designed for the varying quality for flux measurements. This enables a more robust characterization of galaxies.

2) As an improvement of Y3, where asinh magnitudes and colours were used, SOMF uses galaxies' fluxes to group galaxies into SOM cells. While colour is in general more redshift sensitive, at low $S/N$ colour measurements are dominated by noise, leading to large uncertainties. In addition, C3R2 \citep{c3r2_2017, c3r2_2019, c3r2_2021, dc3r2} have found that for galaxies with a fixed colour, there is still a dependence of redshift on magnitude. Flux, on the other hand, is more robust on 'noisy' galaxies with low $S/N$ ratios or even negative flux values. As we move from using asinh colour-magnitude to flux, it is crucial to preserve the asinh magnitude benefit that scales flux logarithmically at high $S/N$ and linearly at low $S/N$. This is achieved by the use of the asinh function in the distance metric equation~(\ref{eqn:distance_metric_expansion}). This modification allows SOMF to retain the benefits of asinh magnitudes, while achieving a more robust measurement of faint galaxies. 

\section{SOMPZ validation using image simulation}\label{app:imsim_nz_validation}

\begin{figure}
    \centering
    \includegraphics[width=\columnwidth]{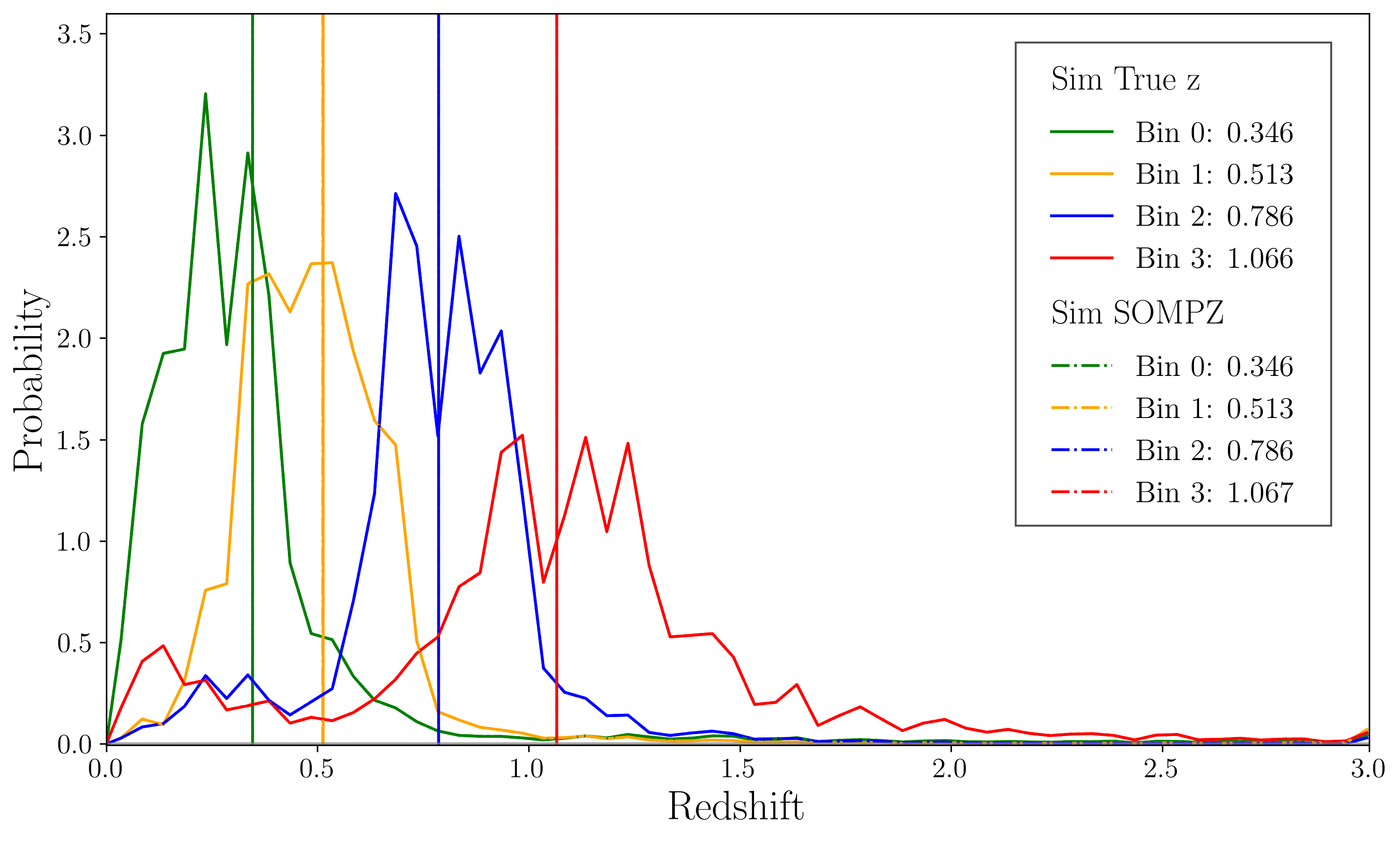}
    \caption{\label{fig:sim_truez_vs_sompz} The redshift distribution calibrated using SOMPZ (dashed) for DES Y6 image simulation galaxies agrees with the true simulated redshift distribution (solid). For the purpose of this depiction, the points at $z=3$ hold the integral of $n(z)$ for $z\ge3$.}
    
\end{figure}

We validate our Y6 SOMPZ algorithm using image simulation with true redshift information (Section \ref{sec:DES_Y6_image_simulations}) in Fig.~\ref{fig:sim_truez_vs_sompz}. The redshift distribution estimated by SOMPZ agrees with the true redshift distribution of the galaxies to $\Delta z < 0.001$ in mean redshift for each tomographic bin, demonstrating the SOMPZ methodology itself has negligible uncertainty. Note that in image simulation we assume perfect calibration samples and transfer function, thus isolating the performance of the SOMPZ method.

\section{Triangular binning}
\label{app:triangular_binning}
The redshift binning in DES Y3 is done using a boxcar function: 
\begin{equation}
n(z) = \sum_i a_i K_i(z)
\end{equation}
\begin{equation}
K_i(z) = \Pi\left( \frac{z}{\Delta_z} - i \right)
\end{equation}
\begin{equation}
\Pi(u) = 
\begin{cases}
1 & \text{if } 0 < u < 1 \\
0 & \text{otherwise},
\end{cases}
\end{equation}
which is equivalent to histogram binning with bin width $\Delta_z$, and we use $\Pi$ to mimic the boxcar shape. However, this histogram binning does not enforce n(z) = 0 at z = 0 and discontinuous at the histogram edges. 

In Y6, we move to a smoother scheme by assuming overlapping triangular bins:
\begin{equation}
K_i(z) = \Lambda\left( \frac{z}{\Delta_z} - i \right)
\end{equation}

\begin{equation}
\Lambda(u) =
\begin{cases}
u & \text{if } 0 < u < 1 \\ 
2 - u & \text{if } 1 < u < 2 \\
0 & \text{otherwise},
\end{cases}
\end{equation}
where we use $\Lambda(u)$ to mimic the triangular shape. Each $\Lambda(u)$ defines a triangular kernel that rises linearly from zero at $z_i - \Delta_z$ to a peak at $z_i$, then falls linearly to zero at $z_i + \Delta_z$. To enforce $n(z) \to 0$ as $z \to 0$, we specifically treat the first bin as an asymmetric triangle that starts at $z = 0$. This ensures both physical and mathematical consistency while preventing divergences in the calculation of observables.

\section{Pileup at z = 3}
\label{app:pileup}
The redshift sample we use contain galaxies with probability with  $p(0 \leq z < 10) > 0$, and we shift all probability with $p(z \geq 3)$ into the final redshift bin at $p(z = 2.99)$. This pileup is done at the last step of analysis during mode compression before sampling inside cosmological inference. There are multiple motivations: 1) The redshift calibration at $z>3$ relies almost entirely on COSMOS. The lack of spectroscopic galaxy limits reliability. 2) The galaxies with $p(z \geq 3)$ constitute less than one per cent in all cases: $[0.29, 0.63, 0.22, 0.34]$. 3) Including more redshift bins increases the computational cost in cosmology inference chains, though this is not the primary reason for the pileup. 

\section{consistency across methods: modes}
\label{app:modes}

\begin{figure}
    \centering
    \includegraphics[width=\columnwidth]{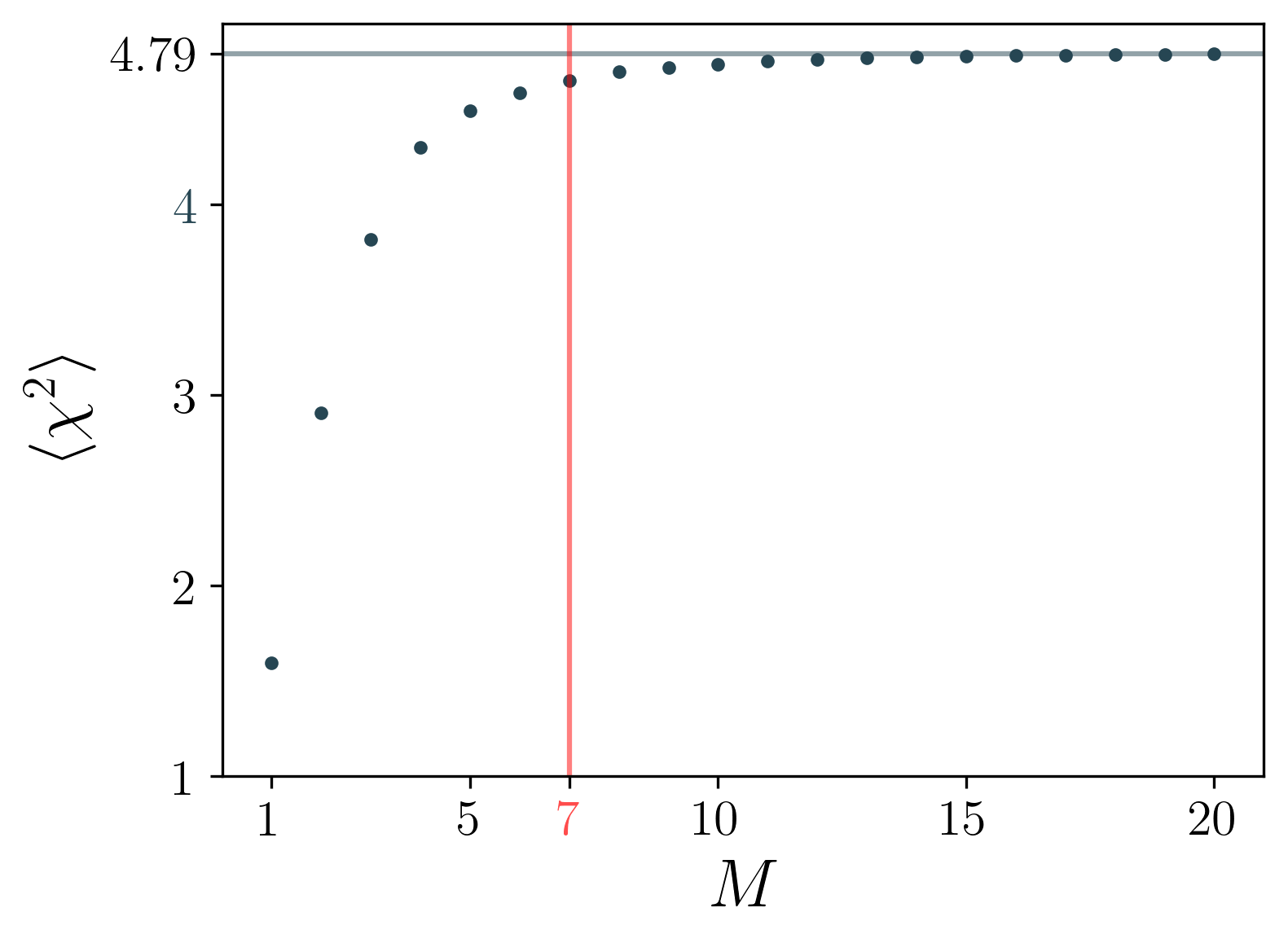}
    \caption{{\label{fig:modes_chi2}
    $\langle \chi^2 \rangle$ distribution as a function of the number of modes $M$ used in mode compression. The horizontal grey line shows the total $\langle \chi^2 \rangle = 4.79$ in our $n(z)$ realizations. The vertical red line indicate the 7 modes we used in the analysis, with the compression loss of $\langle \chi^2 \rangle = 0.14$.
    }
}
\end{figure}

\begin{figure}
    \centering
    \includegraphics[width=\columnwidth]{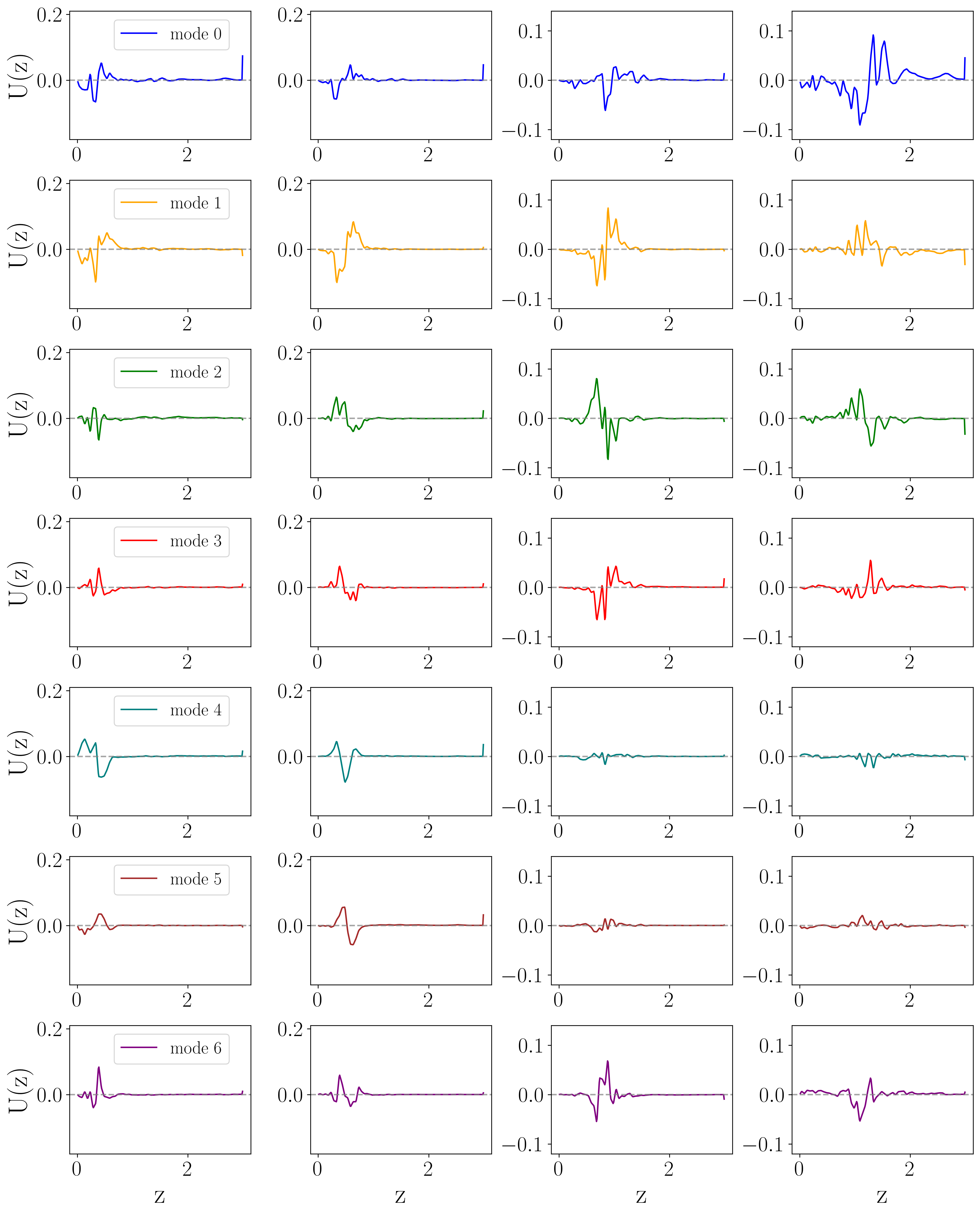}
    \caption{{\label{fig:modes_all_7}
    The basis functions $U_{ij}(z)$ of $8\times10^4$ redshift realizations from SOMPZ + WZ + blending, shown for the four tomographic bins and seven modes. These $U_{ij}(z)$ represent the dimensional reduction of $\delta n(z)$s that impact cosmological inference. The jump in the tail at $z = 3$ arise from the pile up of redshift realizations at $z=3$. 
    }
}
\end{figure}

We analyze the difference between the original redshift realizations and reconstructed redshift realizations after mode compression. The metric $\langle \chi^2 \rangle$ (defined as the average of equation~(\ref{eq:chi2}) over the $n(z)$ samples) is analyzed with varying numbers of retained modes in Fig.~\ref{fig:modes_chi2}. We select $M=7$ modes so that the compression loss is below 0.15. The full 7 modes is plotted in Fig.~\ref{fig:modes_all_7}.

\begin{figure}
    \centering
    \includegraphics[width=\columnwidth]{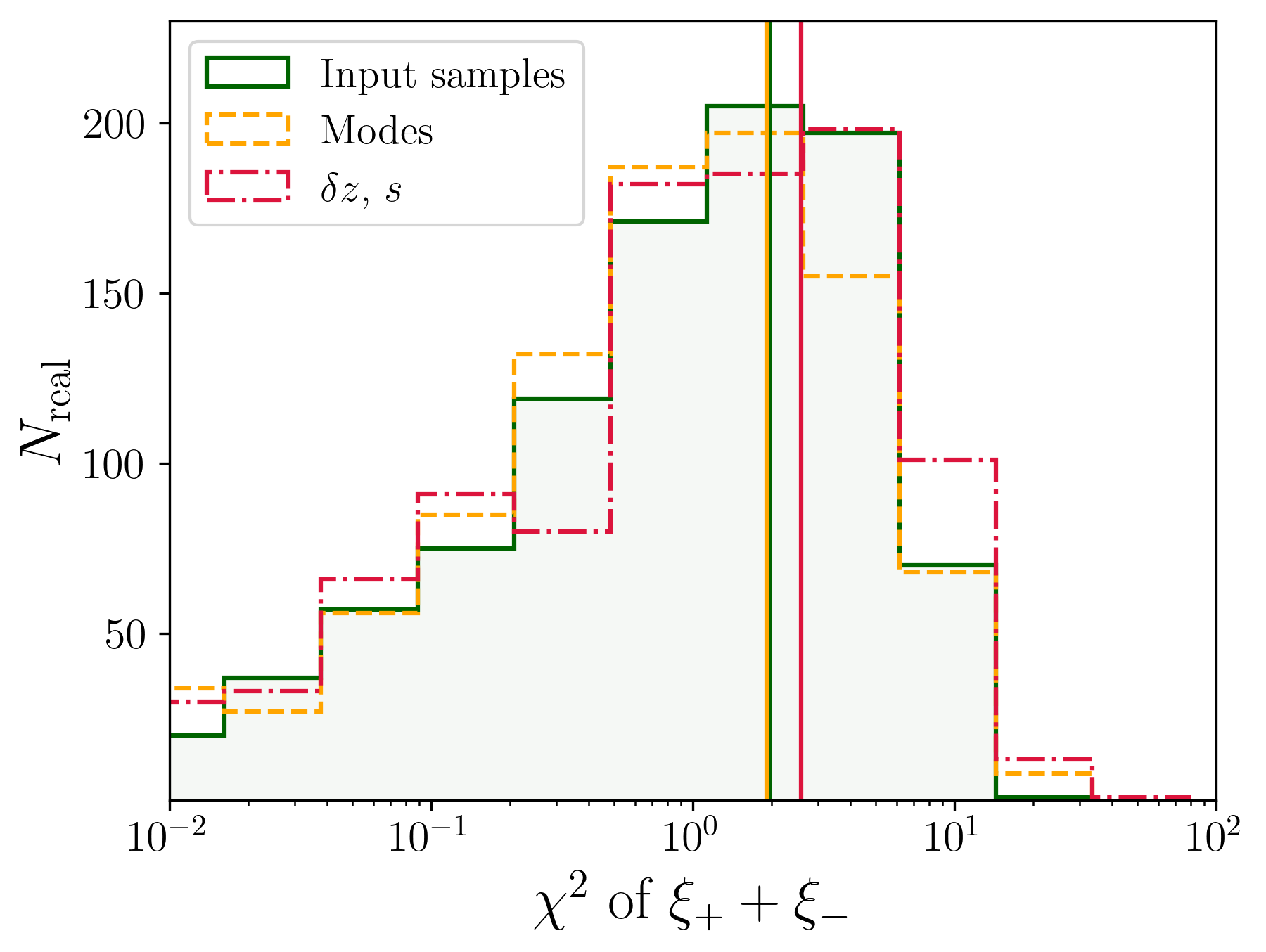}
    \caption{\label{fig:chi2_hist}
    The histogram of the sum of $\xi_+$ and $\xi_-$ $\chi^2$, where each term captures the difference between the datavector generated by a redshift realization and that from the mean redshift distribution. The green, yellow, and red histograms correspond to 1000 $n(z)$ realizations drawn from the original sample, reconstructed after mode compression, and reconstructed from mean-redshift shifts, respectively. The vertical lines show the mean $\chi^2$ values. The reconstructed mode compression closely reproduces the shape and mean of the original redshift realizations' $\chi^2$ distribution.
}
    
\end{figure}

To validate that the mode compression method captures more information from the redshift realizations -- including higher order variations in $\delta n(z)$s and correlation between tomographic bins -- we randomly select 1000 original redshift realizations, 1000 redshift realizations reconstructed from shifts in mean redshift, and 1000 redshift realizations reconstructed from mode compression, and compute their corresponding $\chi^2$ values. Here the $\chi^2$ is computed separately for $\xi_+$ and $\xi_-$, and captures the difference between the datavector generated by a redshift realization and that from the mean redshift distribution. This differs from the earlier definition of $\chi^2$. Fig.~\ref{fig:chi2_hist} shows that the mode compression reproduces both the shape and mean of the $\chi^2$ distribution of original redshift realizations, while the mean redshift shifting method exhibits a larger discrepancy.

\bsp	
\section*{Affiliations}
$^{1}$ Department of Physics, Duke University, Durham, NC 27708, USA\\
$^{2}$ Department of Astrophysical Sciences, Princeton University, Peyton Hall, Princeton, NJ 08544, USA\\
$^{3}$ Department of Physics, Carnegie Mellon University, Pittsburgh, Pennsylvania 15312, USA\\
$^{4}$ Institut de F\'{\i}sica d'Altes Energies (IFAE), The Barcelona Institute of Science and Technology, Campus UAB, 08193 Bellaterra (Barcelona) Spain\\
$^{5}$ Department of Physics and Astronomy, University of Pennsylvania, Philadelphia, PA 19104, USA\\
$^{6}$ Institute of Space Sciences (ICE, CSIC),  Campus UAB, Carrer de Can Magrans, s/n,  08193 Barcelona, Spain\\
$^{7}$ Kavli Institute for Particle Astrophysics \& Cosmology, P. O. Box 2450, Stanford University, Stanford, CA 94305, USA\\
$^{8}$ Department of Physics, Stanford University, 382 Via Pueblo Mall, Stanford, CA 94305, USA\\
$^{9}$ Argonne National Laboratory, 9700 South Cass Avenue, Lemont, IL 60439, USA\\
$^{10}$ Department of Astronomy and Astrophysics, University of Chicago, Chicago, IL 60637, USA\\
$^{11}$ Kavli Institute for Cosmological Physics, University of Chicago, Chicago, IL 60637, USA\\
$^{12}$ University Observatory, LMU Faculty of Physics, Scheinerstr. 1, 81679 Munich, Germany\\
$^{13}$ Fermi National Accelerator Laboratory, P. O. Box 500, Batavia, IL 60510, USA\\
$^{14}$ Nordita, KTH Royal Institute of Technology and Stockholm University, Hannes Alfv\'ens v\"ag 12, SE-10691 Stockholm, Sweden\\
$^{15}$ Institut d'Estudis Espacials de Catalunya (IEEC), 08034 Barcelona, Spain\\
$^{16}$ Physics Department, 2320 Chamberlin Hall, University of Wisconsin-Madison, 1150 University Avenue Madison, WI  53706-1390\\
$^{17}$ SLAC National Accelerator Laboratory, Menlo Park, CA 94025, USA\\
$^{18}$ Department of Applied Mathematics and Theoretical Physics, University of Cambridge, Cambridge CB3 0WA, UK\\
$^{19}$ Department of Physics, University of Genova and INFN, Via Dodecaneso 33, 16146, Genova, Italy\\
$^{20}$ Centro de Investigaciones Energ\'eticas, Medioambientales y Tecnol\'ogicas (CIEMAT), Madrid, Spain\\
$^{21}$ Physik-Institut, University of Zürich, Winterthurerstrasse 190, CH-8057 Zürich\\
$^{22}$ Cerro Tololo Inter-American Observatory, NSF's National\\ Optical-Infrared Astronomy Research Laboratory, Casilla 603, La Serena, Chile\\
$^{23}$ INAF-Osservatorio Astronomico di Trieste, via G. B. Tiepolo 11, I-34143 Trieste, Italy\\
$^{24}$ Department of Physics, University of Michigan, Ann Arbor, MI 48109, USA\\
$^{25}$ Institute of Cosmology and Gravitation, University of Portsmouth, Portsmouth, PO1 3FX, UK\\
$^{26}$ Department of Physics, Northeastern University, Boston, MA 02115, USA\\
$^{27}$ Department of Physics \& Astronomy, University College London, Gower Street, London, WC1E 6BT, UK\\
$^{28}$ School of Mathematics and Physics, University of Queensland,  Brisbane, QLD 4072, Australia\\
$^{29}$ Instituto de Astrofisica de Canarias, E-38205 La Laguna, Tenerife, Spain\\
$^{30}$ Laborat\'orio Interinstitucional de e-Astronomia - LIneA, Av. Pastor Martin Luther King Jr, 126 Del Castilho, Nova Am\'erica Offices, Torre 3000/sala 817 CEP: 20765-000, Brazil\\
$^{31}$ Universidad de La Laguna, Dpto. Astrofísica, E-38206 La Laguna, Tenerife, Spain\\
$^{32}$ Department of Astronomy, University of Illinois at Urbana-Champaign, 1002 W. Green Street, Urbana, IL 61801, USA\\
$^{33}$ Center for Astrophysical Surveys, National Center for Supercomputing Applications, 1205 West Clark St., Urbana, IL 61801, USA\\
$^{34}$ Oxford College of Emory University, Oxford, GA 30054, USA \\
$^{35}$ Jodrell Bank Center for Astrophysics, School of Physics and Astronomy, University of Manchester, Oxford Road, Manchester, M13 9PL, UK\\
$^{36}$ University of Nottingham, School of Physics and Astronomy, Nottingham NG7 2RD, UK\\
$^{37}$ Hamburger Sternwarte, Universit\"{a}t Hamburg, Gojenbergsweg 112, 21029 Hamburg, Germany\\
$^{38}$ Department of Physics, IIT Hyderabad, Kandi, Telangana 502285, India\\
$^{39}$ Universit\'e Grenoble Alpes, CNRS, LPSC-IN2P3, 38000 Grenoble, France\\
$^{40}$ Department of Astronomy/Steward Observatory, University of Arizona, 933 North Cherry Avenue, Tucson, AZ 85721-0065, USA\\
$^{41}$ Jet Propulsion Laboratory, California Institute of Technology, 4800 Oak Grove Dr., Pasadena, CA 91109, USA\\
$^{42}$ Department of Physics and Astronomy, University of Waterloo, 200 University Ave W, Waterloo, ON N2L 3G1, Canada\\
$^{43}$ California Institute of Technology, 1200 East California Blvd, MC 249-17, Pasadena, CA 91125, USA\\
$^{44}$ Instituto de Fisica Teorica UAM/CSIC, Universidad Autonoma de Madrid, 28049 Madrid, Spain\\
$^{45}$ Department of Physics and Astronomy, Pevensey Building, University of Sussex, Brighton, BN1 9QH, UK\\
$^{46}$ Santa Cruz Institute for Particle Physics, Santa Cruz, CA 95064, USA\\
$^{47}$ Center for Cosmology and Astro-Particle Physics, The Ohio State University, Columbus, OH 43210, USA\\
$^{48}$ Department of Physics, The Ohio State University, Columbus, OH 43210, USA\\
$^{49}$ Center for Astrophysics $\vert$ Harvard \& Smithsonian, 60 Garden Street, Cambridge, MA 02138, USA\\
$^{50}$ Lowell Observatory, 1400 Mars Hill Rd, Flagstaff, AZ 86001, USA\\
$^{51}$ Australian Astronomical Optics, Macquarie University, North Ryde, NSW 2113, Australia\\
$^{52}$ George P. and Cynthia Woods Mitchell Institute for Fundamental Physics and Astronomy, and Department of Physics and Astronomy, Texas A\&M University, College Station, TX 77843,  USA\\
$^{53}$ Instituci\'o Catalana de Recerca i Estudis Avan\c{c}ats, E-08010 Barcelona, Spain\\
$^{54}$ Department of Physics, University of Cincinnati, Cincinnati, Ohio 45221, USA\\
$^{55}$ Perimeter Institute for Theoretical Physics, 31 Caroline St. North, Waterloo, ON N2L 2Y5, Canada\\
$^{56}$ Centro de Tecnologia da Informa\c{c}\~ao Renato Archer, Campinas, SP, Brazil - 13069-901\\
$^{57}$ Observat\'orio Nacional, Rua Gal. Jos\'e Cristino 77, Rio de Janeiro, RJ - 20921-400, Brazil\\
$^{58}$ ICTP South American Institute for Fundamental Research\\ Instituto de F\'{\i}sica Te\'orica, Universidade Estadual Paulista, S\~ao Paulo, Brazil",
$^{59}$ Brookhaven National Laboratory, Bldg 510, Upton, NY 11973, USA\\
$^{60}$ Physics Department, Lancaster University, Lancaster, LA1 4YB, UK\\
$^{61}$ Computer Science and Mathematics Division, Oak Ridge National Laboratory, Oak Ridge, TN 37831\\
$^{62}$ School of Physics and Astronomy, University of Southampton,  Southampton, SO17 1BJ, UK\\
\label{lastpage}
\end{document}